\documentclass[aps,prb,twocolumn,superscriptaddress]{revtex4-1}

\usepackage{graphicx}  
\usepackage{dcolumn}   
\usepackage{bm}        
\usepackage{amssymb}   
\usepackage{amsmath}   
\usepackage{hyperref}  

\newcommand*{\BCTO}{Ba$_2$CuTeO$_6$} 						
\newcommand*{\abs}[1]{\lvert{#1}\rvert} 					
\newcommand*{\vect}[1]{\bm{#1}}								
\newcommand*{\matr}[1]{\mathbf{#1}} 						
\newcommand*{\half}[1]{\frac{#1}{2}}  				 		
\newcommand*{\units}{mb\,sr$^{-1}$\,meV$^{-1}$\,Cu$^{-1}$}	

\begin{document}

\title{Spin dynamics of coupled spin ladders near quantum criticality in \BCTO}

\author{David~Macdougal}
\email{david.macdougal@physics.ox.ac.uk}
\affiliation{Clarendon Laboratory, University of Oxford, Parks Road, Oxford, OX1
3PU, United Kingdom}

\author{Alexandra~S.~Gibbs}
\affiliation{ISIS Facility, Rutherford Appleton Laboratory, Harwell Campus,
Didcot, OX11 0QX, United Kingdom}

\author{Tao~Ying}
\affiliation{Institut f{\"u}r Theoretische Festk{\"o}rperphysik, JARA-FIT and
JARA-HPC,
RWTH Aachen University, 52056 Aachen, Germany}
\affiliation{Department of Physics, Harbin Institute of Technology, 150001
Harbin, China}

\author{Stefan~Wessel}
\affiliation{Institut f{\"u}r Theoretische Festk{\"o}rperphysik, JARA-FIT and
JARA-HPC,
RWTH Aachen University, 52056 Aachen, Germany}
\author{Helen~C.~Walker}

\author{David~Voneshen}
\affiliation{ISIS Facility, Rutherford Appleton Laboratory, Harwell Campus,
Didcot, OX11 0QX, United Kingdom}

\author{Fr{\'e}d{\'e}ric~Mila}
\affiliation{Institute of Physics, Ecole Polytechnique F{\'e}d{\'e}rale Lausanne
(EPFL),
1015 Lausanne, Switzerland}

\author{Hidenori~Takagi}
\affiliation{Max Planck Institute for Solid State Research, Heisenbergstrasse 1,
70569 Stuttgart, Germany}

\author{Radu~Coldea}
\affiliation{Clarendon Laboratory, University of Oxford, Parks Road, Oxford, OX1
3PU, United Kingdom}

\date{\today}

\begin{abstract}
We report inelastic neutron scattering measurements of the magnetic excitations
in \BCTO{}, proposed by \textit{ab initio} calculations to magnetically realize
weakly coupled antiferromagnetic two-leg spin-$\half{1}$ ladders. Isolated
ladders are expected to have a singlet ground state protected by a spin gap.
\BCTO{} orders magnetically, but with a small N\'eel temperature relative to the
exchange strength, suggesting that the interladder couplings are relatively
small and only just able to stabilize magnetic order, placing \BCTO{} close in
parameter space to the critical point separating the gapped phase and N\'eel
order. Through comparison of the observed spin dynamics with linear spin wave
theory and quantum Monte Carlo calculations, we propose values for all relevant
intra- and interladder exchange parameters, which place the system on the
ordered side of the phase diagram in proximity to the critical point. We also
compare high field magnetization data with quantum Monte Carlo predictions for
the proposed model of coupled ladders.
\end{abstract}

\maketitle

\section{Introduction}
\label{sec:introduction}

\begin{figure}
\centering
\includegraphics[width=\linewidth,keepaspectratio]{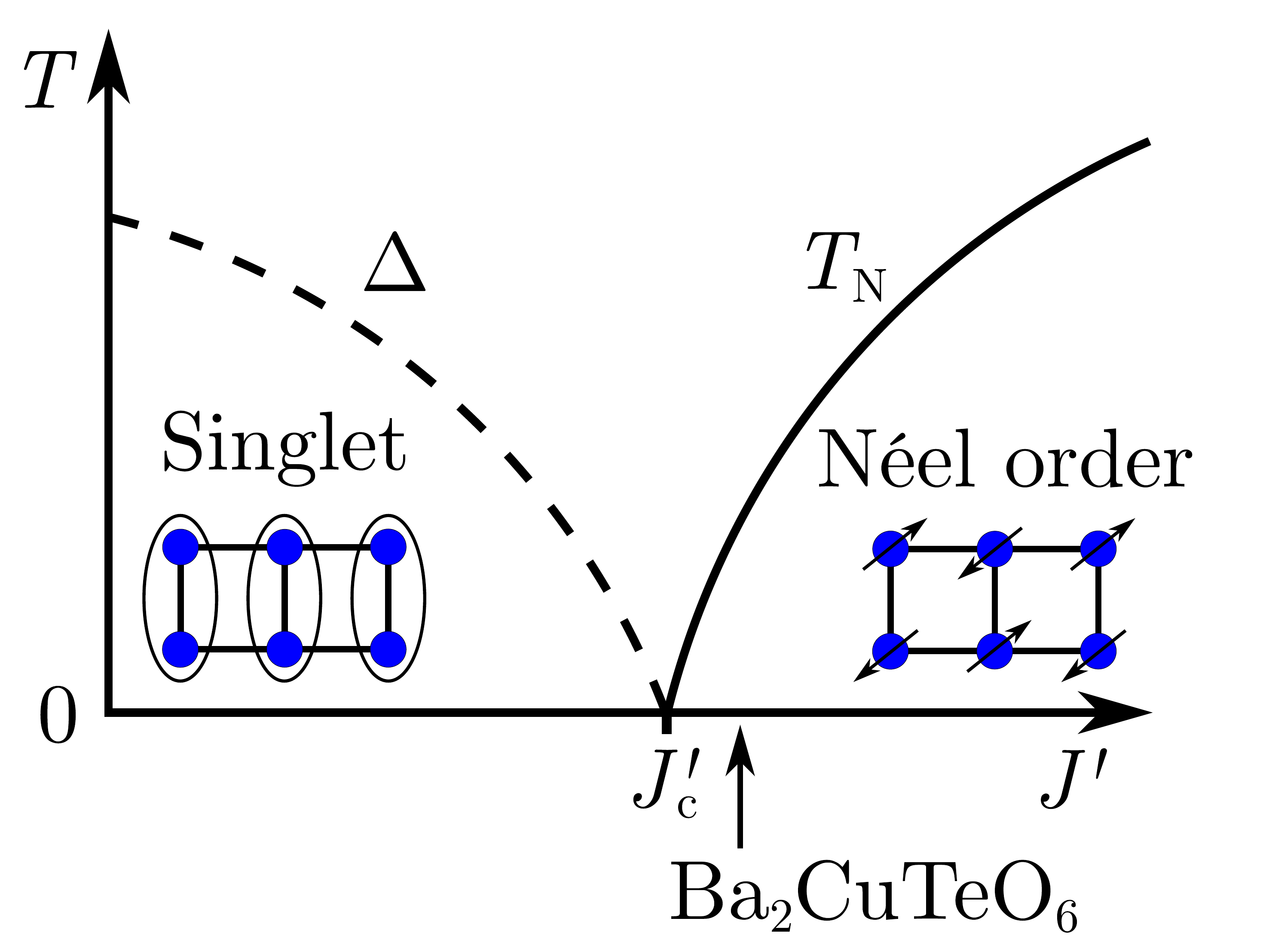}
\caption{(Color online) Schematic phase diagram of two-leg spin-$\half{1}$ AFM
ladders as a function of the in-plane interladder exchange coupling $J'$ [see
Fig.~\ref{fig:Structure}(b)]. Below a critical coupling $J'_{\mathrm{c}}$, the
ground state is an overall singlet with a spin gap $\Delta$. The $J'=0$ state is
schematically illustrated in the inset in the strong rung limit
($J_\mathrm{rung}\gg{}J_\mathrm{leg}$), showing the Cu$^{2+}$ ions (blue
circles) and spin singlet bonds (black ovals). For $J'>J'_{\mathrm{c}}$, AFM
N\'eel order is expected below a finite temperature $T_{\mathrm{N}}$ for nonzero
interplane coupling $J_\mathrm{3D}$. \BCTO{} has been proposed to be located
close to the quantum critical point, on the ordered side of the phase diagram.
\cite{Gibbs2017, Glamazda2017} \label{fig:QPD}}
\end{figure}

Spin ladder systems have attracted considerable interest in the study of high
temperature superconductivity, \cite{Maekawa1996,Uehara1996,Notbohm2007}
Bose-Einstein condensation, \cite{Garlea2007} spinon confinement,
\cite{Lake2010} and Tomonaga-Luttinger liquids
\cite{Klanjsek2008,Thielemann2009,Hong2010} and are known to exhibit interesting
quantum critical behavior. For a two-leg spin-$\half{1}$ antiferromagnetic (AFM)
ladder system, a quantum phase transition is expected to occur as the
interladder exchange coupling $J'$ is varied, as shown in the schematic phase
diagram in Fig.~\ref{fig:QPD}. When $J'$ is lower than a critical value
$J'_\mathrm{c}$, for nonzero leg and rung couplings, the ground state is a
quantum paramagnet with a spin gap $\Delta$ to triplet excitations.
\cite{Barnes1993, Johnston2000, Troyer1997} For larger $J'$ the ground state has
long-range N\'eel order with spin wave excitations.
\cite{Imada1997,Troyer1997,Johnston2000,Furuya2016}
The value of the critical interladder coupling is dependent on
the ratio of the leg and rung couplings and the geometry of the interladder
coupling, with $J'_\mathrm{c}=0.314J$ for in-plane coupled ladders with
$J_\mathrm{leg}=J_\mathrm{rung}=J$.
\cite{Matsumoto2001}

Few candidate systems of coupled spin ladders have been found that are close to
the quantum critical region. (dimethylammonium)(3,5-dimethylpyridinium)CuBr$_4$
has a small N\'eel temperature $T_\mathrm{N}=2$~K relative to the nearly equal
leg and rung couplings $J=7$~K and has a sizable in-plane interladder coupling
$J'=0.32J$, \cite{Hong2014} which suggest that it lies very close to the
critical point on the ordered side of the phase diagram in Fig.~\ref{fig:QPD}.
LaCuO$_{2.5}$ has also been proposed as a possible realization of a nearly
critical system of coupled ladders, \cite{Hiroi1995} with magnetic
ordering observed below $T_\mathrm{N}=125$~K by muon spin rotation
\cite{Kadono1996} ($\mu$SR) and a much larger intraladder coupling $J=1340$~K
extracted from magnetic susceptibility data. \cite{Troyer1997} This
has been supported by tight-binding calculations that predict an interladder
coupling $J'=0.25J$, \cite{Normand1996} close to the critical value
$J'_{\mathrm{c}}=0.115J$ found
from quantum Monte Carlo calculations of the susceptibility for the proposed
three-dimensional (3D) spin ladder system. \cite{Johnston2000}

\begin{figure*}
\centering
\includegraphics[width=\linewidth,keepaspectratio]{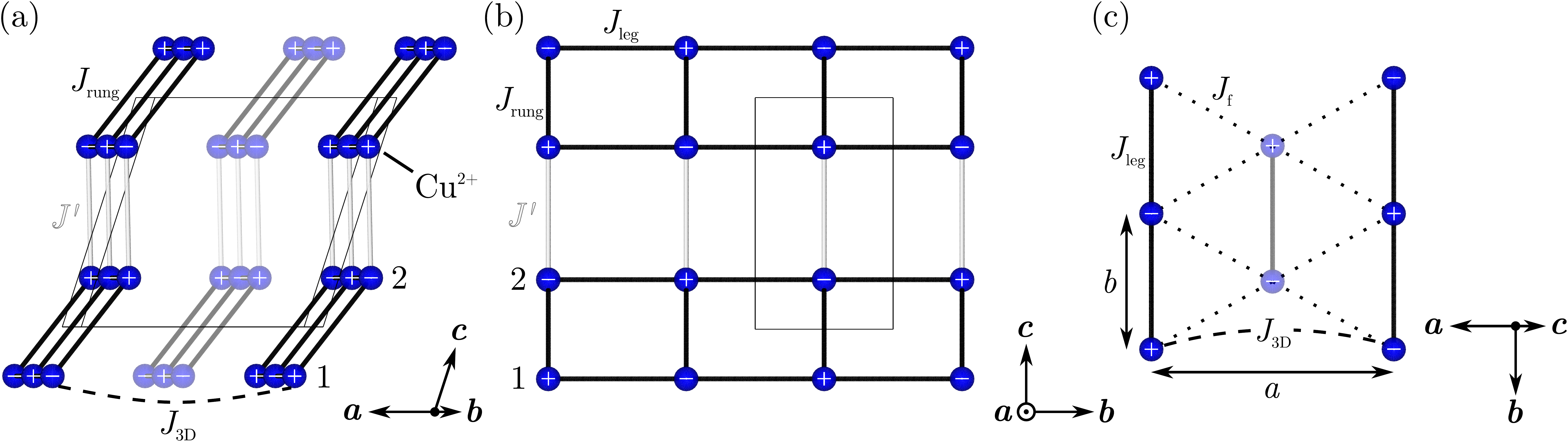}
\caption{(Color online) Crystal structure of \BCTO{} showing Cu$^{2+}$ ions
(blue circles) arranged in two-leg ladders running along $\vect{b}$, with
intraladder exchanges $J_\mathrm{leg}$ and $J_\mathrm{rung}$ (thick black
lines), interladder coupling $J'$ along $\vect{c}$ (gray lines), 
next-nearest-neighbor interplane interaction $J_\mathrm{3D}$ along $\vect{a}$ 
(dashed line), frustrated interplane interaction $J_\mathrm{f}$ in the $ab$ 
plane (dotted lines), the structural monoclinic unit cell (thin black outline),
and the arrangement of up ($+$) and down ($-$) spins in the magnetic ground 
state of the minimal Hamiltonian used in the analysis. The diagrams were 
produced using \textsc{vesta}. \cite{Momma2011} (a) Three buckled planes of 
coupled ladders stacked along $\vect{a}$ with adjacent planes shifted by
$(\vect{a}+\vect{b})/2$. (b) View along $\vect{a}$ with the $c$ axis slightly
into the page, showing a single plane of coupled ladders. (c) Projection of the
structure on the $ab$ plane, showing the interplane interactions $J_\mathrm{f}$
(dotted lines, frustrated) and $J_\mathrm{3D}$ (dashed line,
unfrustrated).\label{fig:Structure}}
\end{figure*}

\BCTO{} crystallizes in an ordered hexagonal perovskite-type structure with
space group $C2/m$ at room temperature. \cite{Kohl1974} It has
been proposed that the Cu$^{2+}$ ions are arranged in weakly coupled two-leg
spin-$\half{1}$ ladders. \cite{Rao2016,Gibbs2017,Glamazda2017}
Figure~\ref{fig:Structure}(b) presents a view of the structure along $\vect{a}$,
showing a single plane of Cu$^{2+}$ ions (blue circles) forming coupled spin
ladders (thick black lines) running along $\vect{b}$. The planes of spin ladders
are stacked along the $a$ axis, as shown in Fig.~\ref{fig:Structure}(a), with
adjacent planes shifted by $(\vect{a}+\vect{b})/2$ relative to one another.
Magnetic susceptibility measurements show an anomaly near 16~K that has
been attributed to a magnetic ordering transition, \cite{Gibbs2017} with $\mu$SR
providing direct evidence for long-range magnetic ordering below
$T_\mathrm{N}=14.1$~K. This temperature is much smaller than the estimated
intraladder exchange strength $J\simeq90$~K. \cite{Gibbs2017, Glamazda2017}
It has been suggested that
the lack of clear signatures of long-range order in NMR, specific heat, and
initial neutron diffraction measurements is evidence of strong quantum
fluctuations and a large suppression of the ordered moment, which would be
expected close to the critical point. \cite{Gibbs2017}

Here we report inelastic neutron scattering (INS) measurements to probe directly
the magnetic excitations in wave vector and energy. We find good agreement
between the observed spin dynamics over the full bandwidth of the excitations
and theoretical predictions for a system of two-leg ladders with sufficiently
strong interladder couplings to stabilize a N\'eel-ordered ground state, and we
propose values for the intra- and interladder couplings consistent with the
observed spin dynamics and previous high field magnetization data.

The rest of this paper is organized as follows. Section~\ref{sec:experiment}
describes the experimental set-up used for the powder INS measurements. The key
features of the dynamics over the full bandwidth of the magnetic excitations are
presented in Sec.~\ref{sec:keyfeatures} and high-resolution measurements of the
low-energy dynamics are reported in Sec.~\ref{sec:lowE}. The following
Sec.~\ref{sec:LSWT} reviews predictions of linear spin wave theory (LSWT) for
two-leg ladders arranged in planes stacked vertically, proposed by \textit{ab
initio} calculations to capture the magnetism of \BCTO{}. Through quantitative
comparison with the observed INS data, values for the intra- and interladder
couplings are extracted, first considering a plane of parallel coupled ladders
(Sec.~\ref{sec:2DLSWT}), which already accounts for most features of the spin
dynamics. One discrepancy is an apparent broadening of the line shape of the
highest-energy excitations and Sec.~\ref{sec:2magnons} proposes a possible
parametrization of this effect. Section~\ref{sec:LSWT_lowE} shows that the
observed suppression of the inelastic magnetic signal at the lowest energies can
be naturally understood as arising from a very weak interaction between parallel
ladder planes; this gives the lowest energy dispersion a 3D
character, in turn leading to a gradual suppression of the spectral weight at
the lowest energies. In Sec.~\ref{sec:QMC}, the key features of the spin
dynamics are compared with quantum Monte Carlo (QMC) calculations for a plane of
coupled ladders and good agreement is found for values of interladder couplings
that are sufficiently strong to put the system on the ordered side of the phase
diagram. Furthermore, similarities and differences between the observed spin
dynamics and that expected for ladders precisely at the critical interladder
coupling strength are discussed in Sec.~\ref{sec:criticalJ'}. As a consistency
check of the overall energy scale of the interactions obtained from comparison
with the LSWT and QMC models, Sec.~\ref{sec:magnetisation} compares experimental
pulsed field magnetization data with mean-field and QMC calculations. Finally,
the conclusions are summarized in Sec.~\ref{sec:conclusion}. The four appendices
contain further technical details of the calculations and analysis:
Appendix~\ref{sec:appendix_LSWT}, LSWT calculation of the dispersion relations 
and INS cross section; Appendix~\ref{sec:appendix_legrung}, its of the LSWT 
model to the measured spin wave spectrum assuming unequal leg and rung 
couplings; Appendix~\ref{sec:appendix_QMC}, QMC calculations; and
Appendix~\ref{sec:appendix_transformation}, transformation between different
crystal structure settings for \BCTO. 

\section{Experimental details}
\label{sec:experiment}

The spin dynamics in a powder sample of \BCTO{} (16~g) was measured using the
direct geometry time-of-flight neutron spectrometer MERLIN at the ISIS neutron
source in the UK. \cite{Bewley2006, MERLIN} The sample used was part of a batch
of polycrystalline material previously found to be single phase using x-ray and
neutron diffraction, with spin susceptibility measurements suggesting less than
$0.1\%$ spin-$\half{1}$ impurities. \cite{Gibbs2017} An incident neutron energy
$E_\mathrm{i}=30$~meV gave an energy resolution on the elastic line of
1.24(2)~meV [full width at half-maximum (FWHM)]. The scale of the magnetic
excitations was found to extend up to $E\simeq16$~meV [see
Fig.~\ref{fig:MERLIN}(a)], so this experimental configuration provided a
suitable energy transfer range with sufficient resolution to probe key features
of the full spectrum. Repetition rate multiplication (RRM) also allowed data to
be collected simultaneously for incident neutrons with $E_\mathrm{i}=12,18,62$,
and 185~meV, although these measurements did not reveal additional features in
the spectrum. A closed cycle refrigerator (CCR) was used to cool the sample to a
base temperature $T=5.8$~K (well below the magnetic ordering transition at
$T_N=14.1$~K) and up to $T=152$~K in the paramagnetic phase. Typical counting
times for each temperature setting were around 9~h at an average proton
current of 152~$\mu$A. The raw neutron counts were converted into absolute
cross section units of \units{} using the measured scattering intensities from a
vanadium standard.

Additional higher-resolution measurements were performed using the direct
geometry time-of-flight neutron spectrometer LET at ISIS. \cite{Bewley2011} INS
data were collected for incident energies $E_\mathrm{i}=1.96$, 3.58, and 21~meV,
with energy resolutions on the elastic line of 0.044(1), 0.105(1), and
1.25(1)~meV (FWHM), respectively, and a CCR was again used to provide
temperature control. Counting times ranged between 7~h at low temperatures
in the magnetically ordered phase to 2.5~h in the paramagnetic phase at high
temperatures, at an average proton current of 40~$\mu$A. The measured integrated
incoherent scattering on the elastic line was used to scale the data collected
at the various incident energies to the same arbitrary units, assuming that the
relative intensity scale factor arises only from the different incident neutron
fluxes. All time-of-flight neutron data were processed using the \textsc{mantid}
data analysis package. \cite{Arnold2014}

\section{Measurements and results}
\label{sec:results}

\subsection{Overview of the spin dynamics}
\label{sec:keyfeatures}

\begin{figure*}
\centering
\includegraphics[width=\linewidth,keepaspectratio]{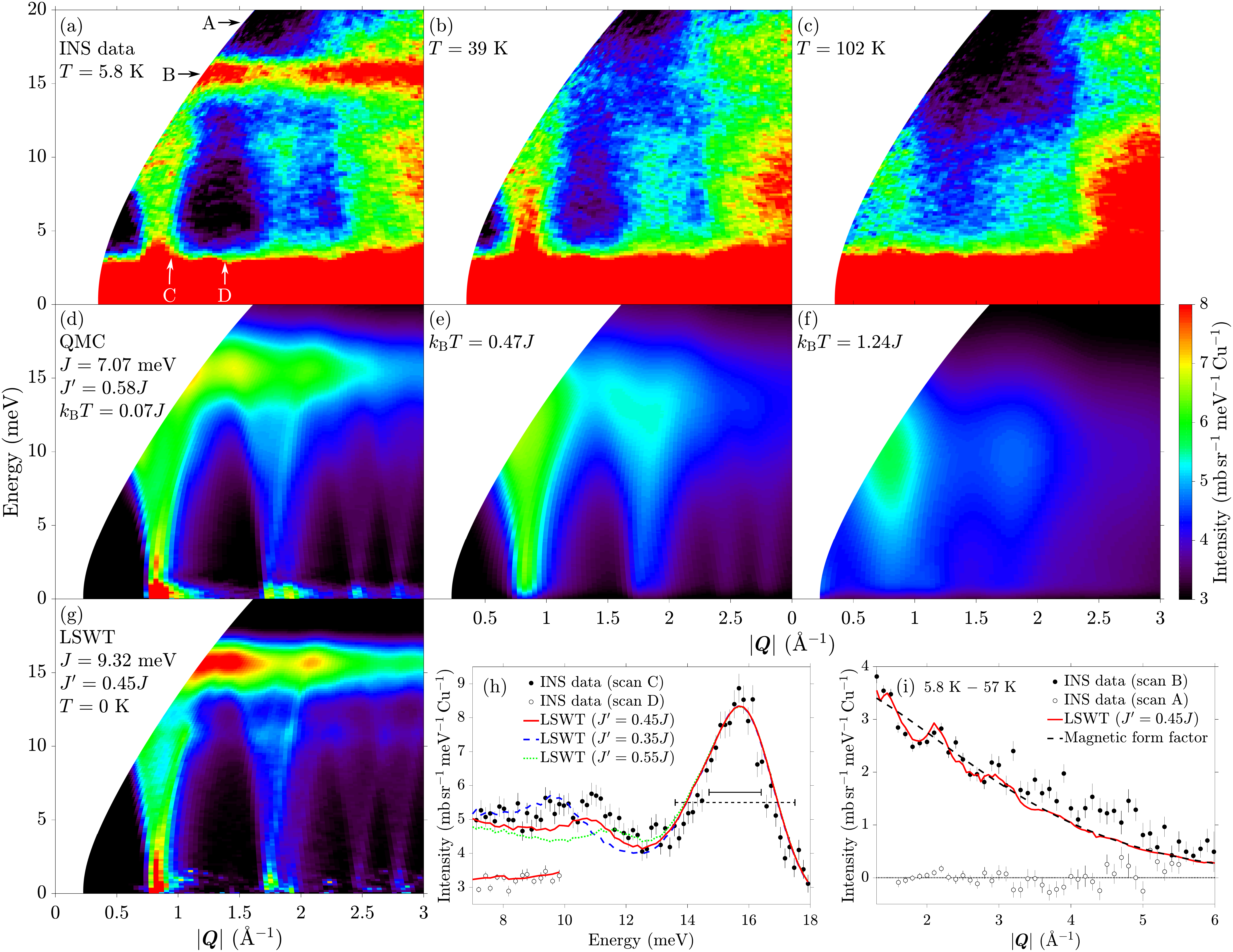}
\caption{(Color online) INS powder data at various temperatures (top row,
MERLIN, $E_\mathrm{i}=30$~meV) compared with QMC (middle row) and LSWT
predictions (bottom row) for no interactions between ladder planes shown in
Fig.~\ref{fig:Structure}(b). The color scale gives the intensities in absolute
units. (h) Energy scan C in (a) (solid symbols), observing a clear intensity dip
in the range 12--14~meV between the V-shaped signal and the higher energy flat
mode. This dip is sensitive to the $J'/J$ ratio, as shown by contrasting the
model prediction for the best-fit value (upper red solid line) with lower/higher
values (dashed blue/dotted green traces) that over-/underestimate the range of
the intensity dip region. The peak centered near 16~meV is broader than expected
based on experimental energy resolution effects (solid horizontal bar),
motivating the assumption of an intrinsic magnon width in the modeling, as
explained in the text. The dashed horizontal bar indicates the FWHM of the flat
mode in the QMC calculation in (d). (i) Wave vector scan B in (a) through the
high-energy flat mode, compared with the LSWT model (solid red line) and the
squared magnetic form factor of Cu$^{2+}$ ions [scaled $f^2(\abs{\vect{Q}})$,
dashed black line]. The higher temperature $T=57$~K data have been subtracted
from the $T=5.8$~K data to remove the nonmagnetic background; open symbols in
(i) indicate the quality of this background subtraction in regions where little
magnetic scattering is expected. All calculations except (i) include a flat
nonmagnetic background contribution, estimated from the measured INS intensities
in regions where the magnetic signal is expected to be small. Scans were
performed along the following directions: A, constant energy $E=[18,21]$~meV;
B, constant energy $E=[14.5,17]$~meV; C, constant scattering angle
$2\theta=[13^\circ,16^\circ]$; and D, constant wave vector
$\abs{\vect{Q}}=[1.3,1.5]$~\AA$^{-1}$.\label{fig:MERLIN}}
\end{figure*}

The powder INS spectrum observed at base temperature is shown in
Fig.~\ref{fig:MERLIN}(a). The key features are a flat `mode' near
$E\simeq16$~meV and a V-shaped dispersive feature centered near
$\abs{\vect{Q}}\simeq0.8$~\AA$^{-1}$. The magnetic character of both inelastic
features is confirmed by their temperature dependence shown in
Figs.~\ref{fig:MERLIN}(b) and \ref{fig:MERLIN}(c); the flat mode has disappeared
at $39$~K, and the V-shaped feature has become overdamped at $102$~K in the 
paramagnetic phase. The flat mode intensity as a function of $\abs{\vect{Q}}$ at
base temperature also follows the squared magnetic form factor of Cu$^{2+}$ ions
[shown in Fig.~\ref{fig:MERLIN}(i)], further confirming its magnetic character. 
The V-shaped scattering is physically attributed to dispersive magnetic 
excitations emanating from a magnetic Bragg peak. The experimentally observed 
V-shaped wave vector magnitude is close to that of the first magnetic Bragg peak
$(h_\mathrm{m}\half{1}1)$ of N\'eel-ordered ladders in each $bc$ plane [shown in
Fig.~\ref{fig:Structure}(b)], with parallel ($h_\mathrm{m}=0$) or alternating
($h_\mathrm{m}=-\half{1}$) stacking between next-nearest-neighbor ladder planes
along $\vect{a}$ being essentially indistinguishable within the resolution of
the present experiment. The intense flat mode is attributed to magnetic
excitations with a high density of states, nondispersive along at least one
crystallographic direction. In the present system, these excitations are magnons
near the maximum of the two-dimensional (2D) dispersion surface for a plane of
coupled ladders, which are nondispersing in the direction normal to the ladder
planes for weak interplane interactions. Note that the strong signal at
$\abs{\vect{Q}}>2.2$~\AA$^{-1}$ in Figs.~\ref{fig:MERLIN}(a)--
\ref{fig:MERLIN}(c) intensifies with increasing temperature, consistent with it 
originating from phonon scattering.

\subsection{Low energy excitations}
\label{sec:lowE}

\begin{figure*}
\centering
\includegraphics[width=\linewidth,keepaspectratio]{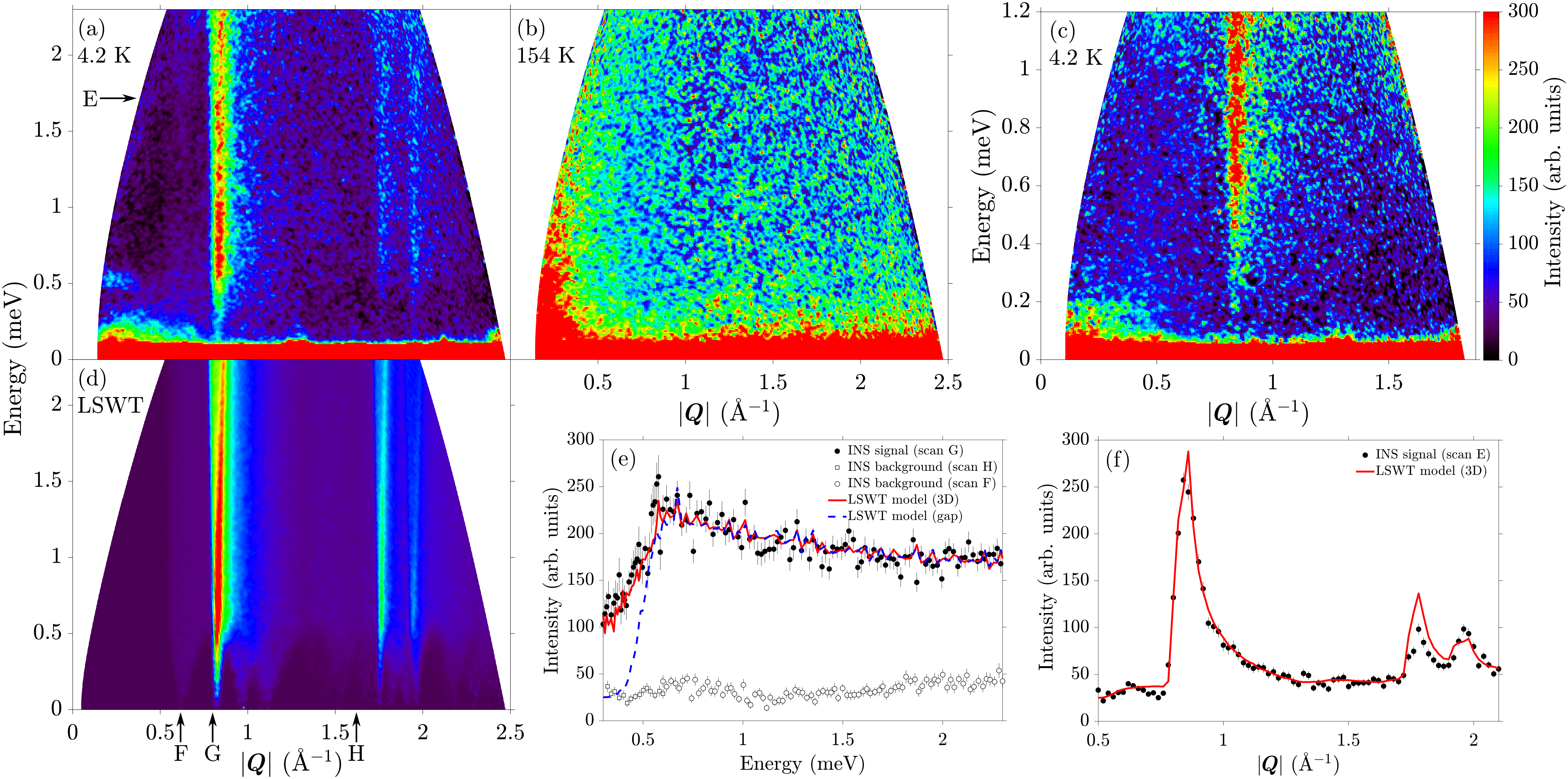}
\caption{(Color online) (a) INS powder data observing a suppression of the
magnetic inelastic signal at low energies compared in (d) with the LSWT
predictions for the case of finite interplane interactions as described in the
text. The gradual intensity suppression at low energies is further confirmed by
higher-resolution data in (c), and the magnetic character of the signal is
established by its disappearance in the paramagnetic phase at high temperature
in (b). (e), (f) Energy and wave vector scans [labeled G in (d) and E in (a),
respectively] compared with the LSWT model with interplane couplings (solid red
lines). Dashed blue line in (e) shows predictions for the alternative finite gap
model described in the text, which does not account for the gradual intensity
suppression below the maximum. Open symbols [scans F and H in (d)] indicate the
nonmagnetic background. Data were collected on LET using $E_\mathrm{i}=3.58$~meV
for (a), (b), (e), and (f), and $E_\mathrm{i}=1.96$~meV for (c). Scans were
performed along the following directions: E, constant energy $E=[1.5,1.9]$~meV;
F, constant wave vector $\abs{\vect{Q}}=[0.56,0.64]$~\AA$^{-1}$; G, constant
wave vector $\abs{\vect{Q}}=[0.6,1.2]$~\AA$^{-1}$; and H, constant wave vector
$\abs{\vect{Q}}=[1.5,1.7]$~\AA$^{-1}$.\label{fig:LET}}
\end{figure*}

High-resolution measurements focusing on the low-energy excitations are
presented in Fig.~\ref{fig:LET}(a). The three narrow V shapes near
$\abs{\vect{Q}}=0.81$, 1.76, and 1.94~\AA$^{-1}$ correspond to regions where
V-shaped dispersive features were also observed in the lower-resolution data in
Fig.~\ref{fig:MERLIN}(a) and are identified with spin wave dispersions coming
out of the magnetic Bragg peaks $(h_\mathrm{m}\half{1}1)$,
$(h_\mathrm{m}\half{3}1)$, and $(h_\mathrm{m}\half{1}3)$, respectively. These
features are again confirmed to be magnetic as they are not present at high $T$
in Fig.~\ref{fig:LET}(b) ($T=154$~K). Focusing on the low-energy region one can
observe a clear intensity decrease upon decreasing energy below
$\simeq0.55$~meV, see Figs.~\ref{fig:LET}(a) and \ref{fig:LET}(e) 
(solid points). This is a gradual decrease, rather than a sharp cut-off, with a
clear inelastic signal observed down to the lowest resolvable energies. 
Figure~\ref{fig:LET}(c) presents even higher-resolution INS data, which show 
that a spin gap, if present, is smaller than an upper bound of $\simeq0.15$~meV.

\section{Analysis}
\label{sec:analysis}

\subsection{Linear spin wave theory for coupled ladders}
\label{sec:LSWT}
To parametrize the INS data, we used linear spin wave theory (LSWT) for a system
of parallel two-leg spin-$\half{1}$ ladders arranged in planes with monoclinic
stacking, as shown in Fig.~\ref{fig:Structure}(a). We assume Heisenberg AFM
exchanges for all intraladder ($J_\mathrm{rung}$ and $J_\mathrm{leg}$) and
interladder couplings ($J'$ between adjacent ladders in the $bc$ plane and
$J_\mathrm{3D}$ between parallel ladder planes shifted by $\vect{a}$),
neglecting the frustrated interaction $J_\mathrm{f}$ between offset planes [see
Fig.~\ref{fig:Structure}(c)] as its effects cancel at the mean-field level. In
the mean-field ground state, the spins are AFM aligned in the ladder planes, and
parallel planes displaced by $\vect{a}$ are oppositely aligned, as shown by the
alternation of $+$ and $-$ signs in Fig.~\ref{fig:Structure} (dark blue sites).
There are two distinct spin wave branches with dispersion relations obtained
(see Appendix~\ref{sec:appendix_LSWT} for details) as
\begin{equation}
\hbar\omega_{\vect{Q}}^{\pm}=\sqrt{A^2-(C\mp{}D_0)^2},
\label{eq:Dispersion}
\end{equation}
where the upper (lower) label corresponds to excitations with even (odd) parity
with respect to swapping sites 1 and 2 in the primitive cell. Each dispersion
branch is doubly degenerate, corresponding to magnons with spin component
$S_z=\pm1$, where $z$ defines the direction of the ordered spins in the ground
state. The parameters determining the dispersion relations at a general
wave vector $\vect{Q}=h\vect{a}^\ast+k\vect{b}^\ast+l\vect{c}^\ast$ [expressed 
as $(h,k,l)$ in reciprocal lattice units of the monoclinic $C2/m$ structural 
cell] are
\begin{align}
\begin{split}
A&=2S\left(J_\mathrm{leg}+\frac{J_\mathrm{rung}}{2}
+\frac{J'}{2}+J_\mathrm{3D}\right)\\
C&=-2S(J_\mathrm{leg}\cos{\vect{Q}\cdot\vect{b}}
+J_\mathrm{3D}\cos{\vect{Q}\cdot\vect{a}})\\
D&=S\left(J_\mathrm{rung}+J'e^{-2\pi{}il}\right)e^{2\pi{}i(h\xi+l\zeta)}\\
&\equiv{}D_0e^{i\phi}, \quad D_0=\abs{D}.
\end{split}
\label{eq:ACD}
\end{align}
Here $\xi=-0.1866(1)$ and $\zeta=0.4299(1)$ define the separation between the
two spins on each rung
$\vect{r}_2-\vect{r}_1=\xi\vect{a}+\zeta\vect{c}$,\cite{Gibbs2017} where the
subscripts 1 and 2 refer to the numbered positions in
Figs.~\ref{fig:Structure}(a) and \ref{fig:Structure}(b).

\subsubsection{A plane of parallel coupled ladders}
\label{sec:2DLSWT}
\begin{figure*}
\centering
\includegraphics[width=\linewidth,keepaspectratio]{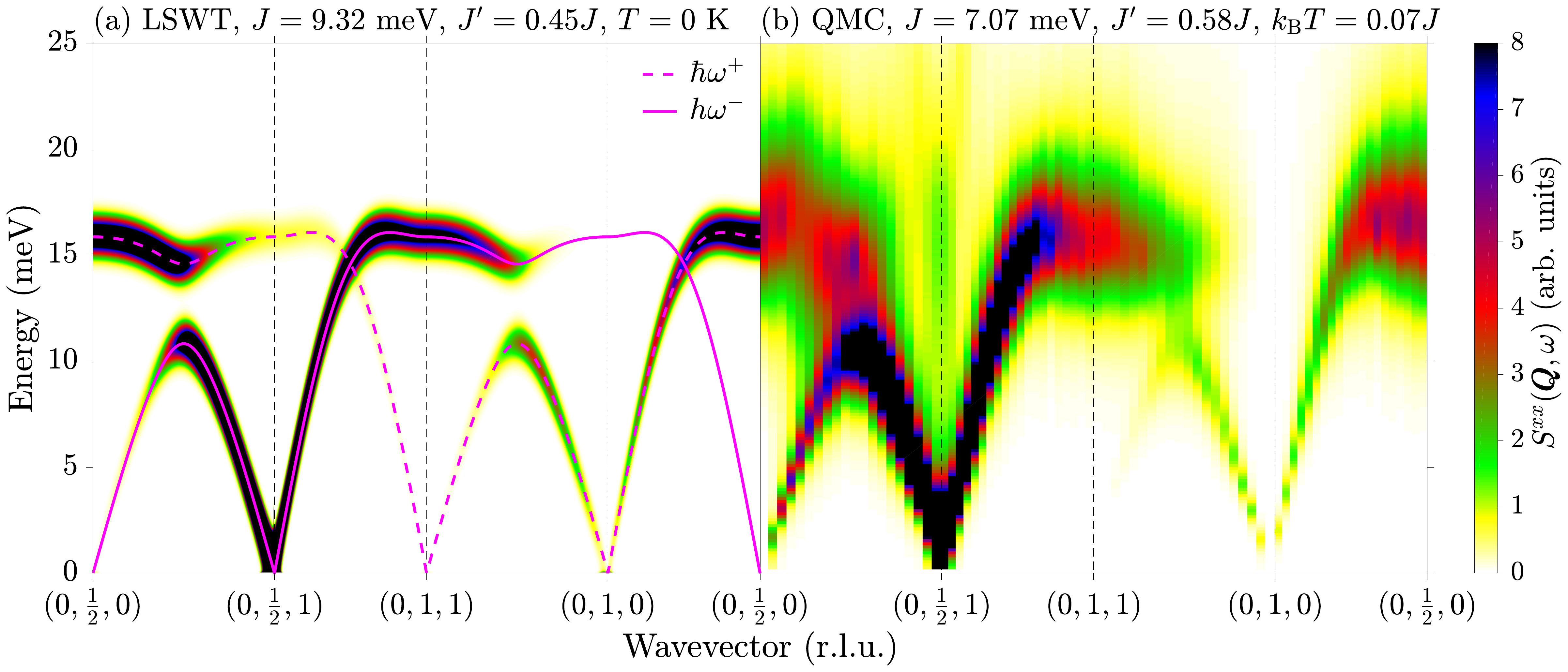}
\caption{(Color online) Color map representation of the dynamical structure
factor $S^{xx}(\vect{Q},\omega)$ along high symmetry directions in reciprocal
space for a plane of coupled ladders [shown in Fig.~\ref{fig:Structure}(b)] with
intraladder coupling $J_\mathrm{rung}=J_\mathrm{leg}=J$ and interladder coupling
$J'$. (a) LSWT spectrum using Eq.~(\ref{eq:Sxx}) in
Appendix~\ref{sec:appendix_LSWT} with a constant $\sigma=0.53$~meV to match the
observed FWHM of the elastic line in the INS MERLIN data. Solid and dashed lines
indicate the magnon dispersion relations given in Eq.~(\ref{eq:Dispersion}). (b)
QMC spectrum at temperature $k_\mathrm{B}T=0.07J$. The exchange parameters used
in each case are those corresponding to the best fit of each model to the powder
INS data in Fig.~\ref{fig:MERLIN}(a). Wave vectors $(h,k,l)$ are given in
reciprocal lattice units (r.l.u.) of the structural monoclinic unit cell in
Fig.~\ref{fig:Structure}, and the color scale is in arbitrary
units.\label{fig:LSWT&QMC}}
\end{figure*}

We first consider the limit of no interactions between spins in different ladder
planes ($J_\mathrm{3D}=0$), for which there is no dispersion along $h$.
Figure~\ref{fig:LSWT&QMC}(a) plots the corresponding spin wave dispersion
relations along a path of high symmetry directions in reciprocal space for the
case of equal rung and leg couplings ($J_\mathrm{rung}=J_\mathrm{leg}=J$) and
finite interladder exchange $J'$. The $\hbar\omega^-$ dispersion surface is
gapless at the $(0,\half{1},1)$ wave vector corresponding to in-plane N\'eel
order and disperses along both in-plane directions forming a linearly
dispersing, elliptical cone at low energies, where the dispersion along
$\vect{b}^\ast$ is due to $J_\mathrm{leg}$ and the dispersion along
$\vect{c}^\ast$ is due to the interladder coupling $J'$. Note that magnons at
the highest energies have extended regions with very little dispersion along
$\vect{b}^\ast$ or $\vect{c}^\ast$, which would lead to a large density of
states at those energies upon spherical averaging of the spectrum.

\begin{figure*}
\centering
\includegraphics[width=\linewidth,keepaspectratio]{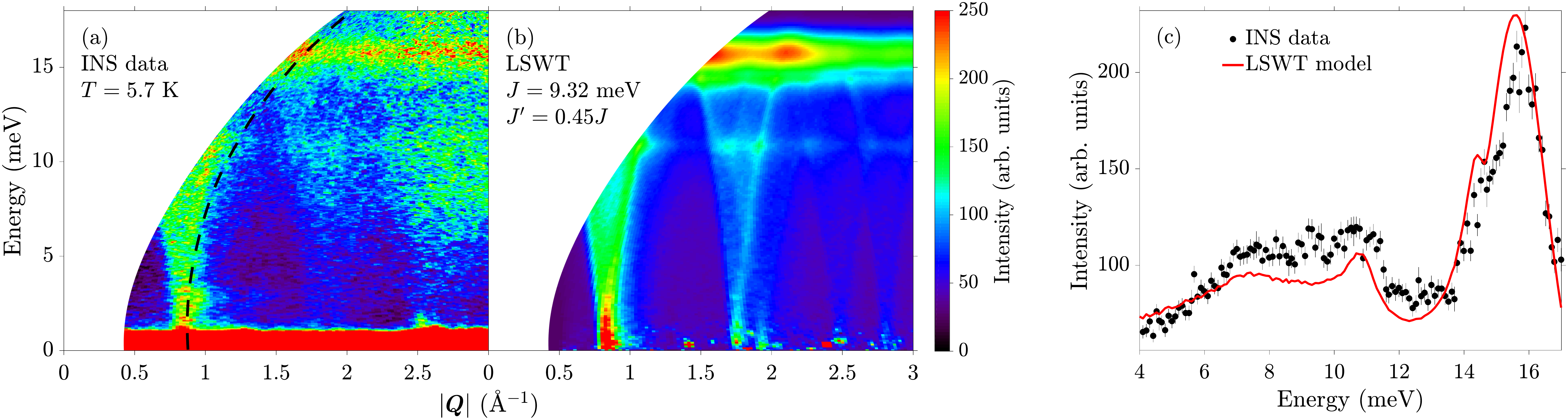}
\caption{(Color online) (a) Additional INS powder data (LET,
$E_\mathrm{i}=21$~meV, $T=5.7$~K) compared in (b) with the LSWT model for
decoupled ladder planes [same exchange parameters as in
Fig.~\ref{fig:MERLIN}(g)]. The color scale gives the intensities in arbitrary
units. (c) Energy scan at constant scattering angle
($2\theta=[7^\circ,25^\circ]$) through the
INS data (black circles) and the LSWT calculation (red line), probing the
V-shaped feature, the intensity dip region, and the high-energy
mode. The center of the scan is indicated by the dashed line in (a).
\label{fig:LET_21meV}}
\end{figure*}

To compare with the experimental INS powder data, we perform a spherical average
of the LSWT one-magnon spectrum, including the full wave vector and energy
dependence of the one-magnon neutron scattering cross section, the neutron
polarization factor, the Bose temperature factor, the spherical magnetic form
factor for Cu$^{2+}$ ions, and the convolution with the estimated instrumental
resolution (see Appendix~\ref{sec:appendix_LSWT} for details). The best fit to
the data assuming equal leg and rung couplings
($J_\mathrm{leg}=J_\mathrm{rung}=J$), as predicted by \textit{ab initio}
calculations, \cite{Gibbs2017} is shown in Fig.~\ref{fig:MERLIN}(g) (a
more refined analysis for $J_\mathrm{leg}\neq J_\mathrm{rung}$ will be presented
in Appendix~\ref{sec:appendix_legrung}). In the fit, both the intra- and
interladder exchanges $J$ and $J'$ were varied, as well as an overall intensity
scale factor, and the best-fit parameter values are listed in
Fig.~\ref{fig:MERLIN}(g). As explained before, the region above
$\abs{\vect{Q}}\simeq2.2$~\AA$^{-1}$ is dominated by phonon scattering and will
not be discussed further.
The model reproduces well the key features of the observed magnetic inelastic
signal in Fig.~\ref{fig:MERLIN}(a); in particular, the primary V-shaped feature
is attributed to the $\hbar\omega^-$ spin wave cone emanating from the magnetic
Bragg rod at $(h,\half{1},1)$, and the flat mode near $E\simeq16$~meV is
attributed to the almost dispersionless magnons near the top of the 2D
dispersion in Fig.~\ref{fig:LSWT&QMC}(a), which are also dispersionless in the
third direction for decoupled ladder planes. The feature in the INS data in
Fig.~\ref{fig:MERLIN}(a) that is most sensitive to the strength of the
interladder coupling $J'$ is an apparent narrowing of the V-shaped signal before
it merges with the higher-energy flat mode, with a clear intensity dip in this
intermediate energy region. This is most clearly seen in the energy scan in
Fig.~\ref{fig:MERLIN}(h), which reveals an intensity dip in the range 12--14~meV
between the top of the V shape and the flat mode at higher energies. The energy
dependence of the intensity in this scan is best described for $J'=0.45J$ (upper
red solid line); for weaker couplings the intensity dip region is significantly
wider than observed, as illustrated for $J'=0.35J$ (blue dashed line), whereas
for stronger couplings the dip region narrows and becomes less distinct, as
illustrated for $J'=0.55J$ (green dotted line). For ease of comparison, the
value of the intraladder coupling $J$ was adjusted to keep the energy of the
higher-energy flat mode unchanged in all the above cases. Additional INS
measurements, shown in Fig.~\ref{fig:LET_21meV}(a) with the LSWT calculation in
Fig.~\ref{fig:LET_21meV}(b) and an energy scan in Fig.~\ref{fig:LET_21meV}(c), 
provide further confirmation that the LSWT model for decoupled ladder planes 
captures well the key features of the magnetic inelastic response.

\subsubsection{Width of high energy flat mode}
\label{sec:2magnons}

\begin{figure}
\centering
\includegraphics[width=\linewidth,keepaspectratio]{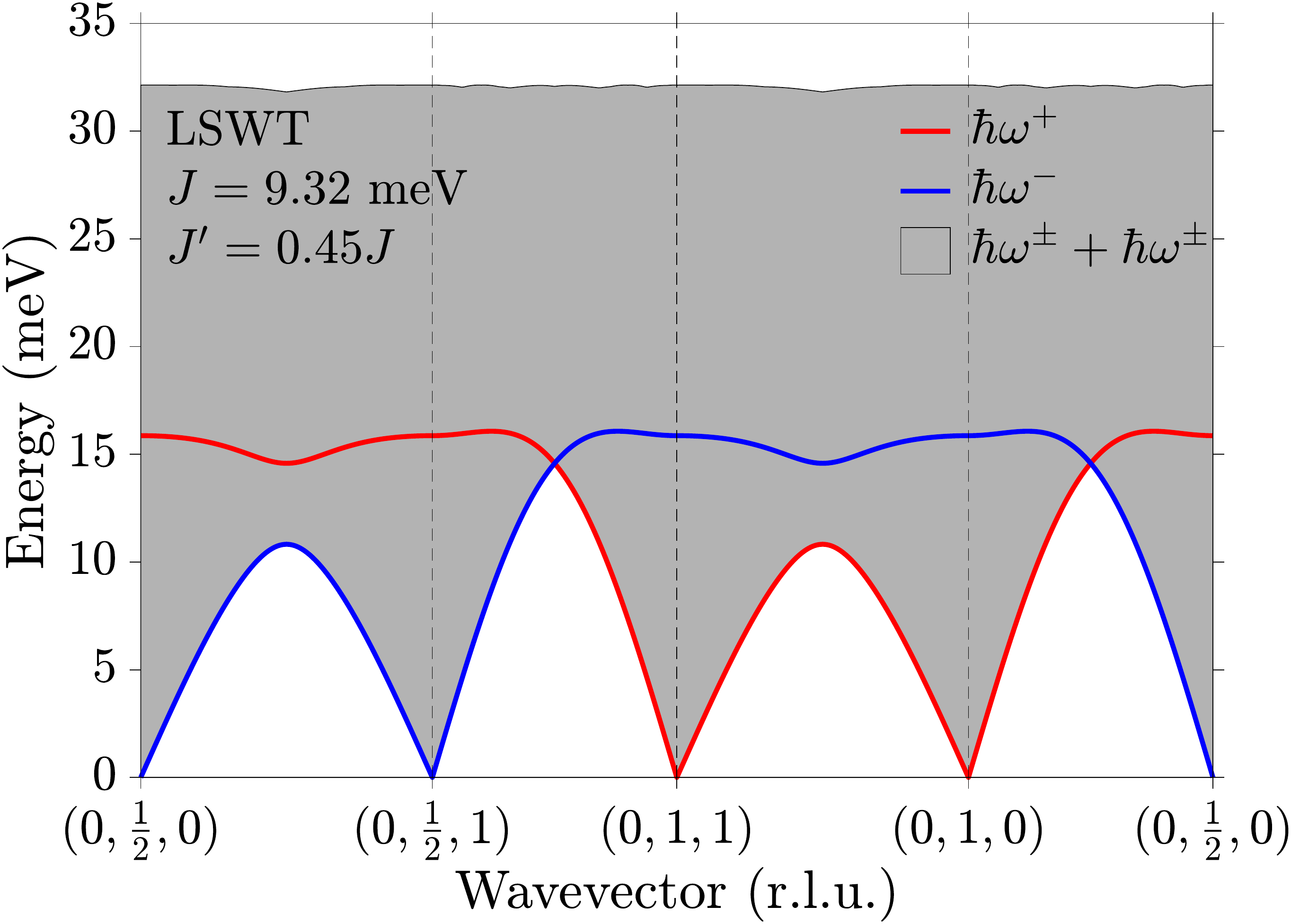}
\caption{(Color online) Phase space for two-magnon excitations (gray shaded
area), calculated using LSWT for isolated ladder planes. The one-magnon
dispersions [red and blue solid lines, Eq.~(\ref{eq:Dispersion})] overlap with
the continuum above an energy threshold. The exchange parameters are as in
Fig.~\ref{fig:LSWT&QMC}(a).\label{fig:LSWT_2magnons}}
\end{figure}

A notable feature of the energy scan in Fig.~\ref{fig:MERLIN}(h) is that the
strong peak centered around $E\simeq16$~meV appears broader than expected based
on the estimated instrumental resolution and powder averaging (combined expected
FWHM represented by the solid horizontal bar at the peak's half-maximum), and
this observed broadening cannot be accounted for by using moderately different
leg and rung couplings. The width of the flat mode and its relative intensity
compared to the V-shaped feature are best reproduced [solid red line in
Fig.~\ref{fig:MERLIN}(h)] if a Gaussian broadening (of empirical FWHM 2.1~meV)
is assumed for high-energy magnons. Finite one-magnon lifetime effects are
generally associated with one$\rightarrow$two-magnon decay processes and
Fig.~\ref{fig:LSWT_2magnons} illustrates the phase space (shaded area) for
two-magnon processes; the one-magnon dispersions overlap with this region above
an energy threshold of 14.5~meV, so in all comparisons with the LSWT model we
have assumed an intrinsic lifetime for magnons above this energy threshold and
this seems to provide a good empirical parametrization of the data.

\subsubsection{Low energy excitations and couplings between ladders planes}
\label{sec:LSWT_lowE}
The observed nonmonotonic energy dependence of the magnetic intensity at low
energies, with a gradual suppression of the signal below $\simeq$0.55~meV [shown
in Figs.~\ref{fig:LET}(a) and \ref{fig:LET}(e)], cannot be explained by the 
above model of isolated ladder planes with Heisenberg exchanges. This model 
would predict a gapless, elliptical spin wave cone at the lowest energies with 
a constant spectral density of states in a wide energy range above zero energy.
To identify
the possible origin of the observed loss of spectral weight at low energies, we
have separately considered two possible extensions of the previous model: (i)
addition of a finite interaction between parallel ladder planes, which gives the
lowest-energy magnons a 3D dispersion with a suppressed density of states, and
(ii) presence of a finite spin gap, still assuming isolated ladder planes. Such
a spin gap may physically originate from Dzyaloshinskii-Moriya couplings
(symmetry allowed as the Cu-Cu bonds are not centrosymmetric) or other
anisotropic exchanges.

We first considered the case of a finite AFM interaction $J_\mathrm{3D}$ between
parallel ladder planes displaced along $\vect{a}$, for which the LSWT dispersion
relations are given in Eq.~(\ref{eq:Dispersion}). (We have also numerically
calculated the spherically averaged spin wave spectrum for ferromagnetic
$J_\mathrm{3D}$, and the results are rather similar, with only small differences
at the very lowest energies; the data are not sensitive enough to distinguish
between these two scenarios, so in the following we consider just the AFM
$J_\mathrm{3D}$ case for concreteness.) The spherically averaged spin wave
spectrum for the best-fit value $J_\mathrm{3D}\simeq5\times10^{-4}J$, with $J$
and $J'$ fixed at the values determined previously, is shown in
Fig.~\ref{fig:LET}(d) and compares well with the INS data in
Fig.~\ref{fig:LET}(a). In particular, the model reproduces well the positions
and relative intensities of the three visible V-shaped features and their
intensity drop-off at low energies. This is more clearly seen in the energy scan
through the center of the primary V-shaped feature in Fig.~\ref{fig:LET}(e),
where the observed intensity profile is well captured by the model calculation
(solid red line). The only fitted parameters for this energy scan are
$J_\mathrm{3D}$ and an overall intensity scale factor. In this model, the
nonmonotonic intensity profile arises from the cross-over from a 3D to a 2D
spectral density of states upon increasing energy above the interplane zone
boundary energy $E_\mathrm{3D}=
2S\sqrt{2J_\mathrm{3D}(2J_\mathrm{leg}+J_\mathrm{rung}+J')}=0.54$~meV
[which occurs at $(0,\half{1},1)$]. Below this energy, the magnons have a 3D
dispersion, linear in all directions at the lowest energies, for which the
spectral density of states has an $\hbar\omega$ dependence [as the density of
states varies as $(\hbar\omega)^2$ and the dynamical structure factor for AFM
magnons varies as $1/\hbar\omega$]. Above the saddle point energy
$E_\mathrm{3D}$, the magnon dispersion is predominantly in the plane of the
ladders and the spectral density of states is constant for linearly dispersing
magnons in 2D (as the density of states varies as $\hbar\omega$ and the
structure factor as $1/\hbar\omega$), explaining the near constant intensity at
these energies. The shape of the intensity profile in wave vector scans at
energies above $E_\mathrm{3D}$ in Fig.~\ref{fig:LET}(f), showing an asymmetric
tail towards higher wave vectors, is also well explained by quasi-2D linearly
dispersing magnons (red line).

In the alternative parametrization of isolated ladder planes ($J_\mathrm{3D}=0$)
with a finite gap $\Delta$, we assume the dispersion relations have the modified
form
\begin{equation}
\hbar\widetilde{\omega}_{\vect{Q}}^{\pm}=
\sqrt{(\hbar\omega_{\vect{Q}}^{\pm})^2+\Delta^2},
\label{eq:Spin_gap}
\end{equation}
with the dynamical correlations as given in Eq.~(\ref{eq:Sxx}), but with
$\hbar\omega_{\vect{Q}}$ replaced by $\hbar\widetilde{\omega}_{\vect{Q}}$.
The dashed blue line in Fig.~\ref{fig:LET}(e) shows the calculation for
$\Delta=0.5$~meV. Such a model clearly predicts a much more abrupt intensity
drop-off than is actually observed. We note that our analysis does not preclude
the existence of a much smaller spin gap with an upper bound of
$\simeq0.15$~meV, the resolution of our experiments (a spin gap of order 0.2~meV
may be expected since measurements in applied field show a transition near a
critical field of 15~kOe, associated with a spin-flop transition).\cite{Rao2016}
A similar intensity suppression has been reported in a spin chain material, in
which doping with nonmagnetic impurities opens a pseudogap. \cite{Simutis2013}
In that case, the measured intensity decreases monotonically at low energies,
whereas we observe a nonmonotonic dependence with a peak at $\simeq0.55$~meV,
which is better described by a saddle point in the dispersion at the interplane
zone boundary. Furthermore, the pseudogap was reported for a 1\% doping and
there is estimated to be only 0.1\% nonmagnetic impurities in the present
system. \cite{Gibbs2017}
Based on the above two parametrizations, we therefore conclude that the finite
interplane couplings are the most likely origin for the observed signal
suppression at low energies, and we attribute the intensity maximum near
0.55~meV with the interplane magnetic zone boundary energy.

\subsection{Quantum Monte Carlo calculations for a plane of coupled ladders}
\label{sec:QMC}
The LSWT description used so far in the analysis relies on the assumption that
the system is located deep in the ordered side of the schematic phase diagram in
Fig.~\ref{fig:QPD}, where zero-point quantum fluctuations are relatively small.
However, this is not the case for values of the interladder coupling $J'$ that
put the system still in the ordered phase, but close to the critical point
separating it from the gapped singlet phase at low $J'$. Since the estimated
$J'/J$ value is close to the expected critical threshold for the onset of
magnetic order, it is insightful to compare the INS results with more elaborate
theories that better capture quantum fluctuation effects. For this purpose, we
have performed quantum Monte Carlo (QMC) calculations for a plane of coupled
ladders with intraladder coupling $J_\mathrm{leg}=J_\mathrm{rung}=J$ and
interladder coupling $J'$. The QMC simulations were performed using the
stochastic series expansion method with directed loop updates,
\cite{Sandvik1999, Syljuasen2002, Alet2005} using an efficient scheme to measure
imaginary time displaced spin-spin correlation functions,
\cite{Michel2007a,*Michel2007} and the stochastic analytic continuation method
in the formulation of Ref.~\onlinecite{Beach2004} to obtain the dynamical spin
structure factor (see Appendix~\ref{sec:appendix_QMC} for details). The obtained
dynamical correlations along high symmetry directions in reciprocal space are
shown in Fig.~\ref{fig:LSWT&QMC}(b) for the best-fit parameter values as listed
in the figure title. The calculation qualitatively resembles many features
already captured at the LSWT level [Fig.~\ref{fig:LSWT&QMC}(a)], such as how the
strongest scattering weight disperses in the Brillouin zone, the spin wave cone 
emerging out of the N\'eel order Bragg peak, and the extended regions with 
little dispersion at the top of the excitation bandwidth. The dynamical 
correlations
appear significantly broadened, in particular at high energies, and this effect
may be interpreted as being partly due to multimagnon contributions (see
Appendix~\ref{sec:appendix_QMC}). This intrinsic broadening is illustrated by
the horizontal dashed bar plotted near the main peak's half-maximum in
Fig.~\ref{fig:MERLIN}(h); the width is of a comparable extent to the line shape
width observed experimentally.

The best-fit values for the intra- and interladder exchanges were again obtained
by fitting the calculated spherically averaged dynamical correlations (including
all the relevant neutron scattering intensity prefactors and instrumental
resolution effects) to the measured low-temperature magnetic INS signal, with
the aim of reproducing the high-energy flat mode and the V-shaped signal. The
overall intensity scale factor was determined by fitting to a cut through the
high-energy flat mode. The best-fit result is shown in Fig.~\ref{fig:MERLIN}(d)
for comparison with the data in Fig.~\ref{fig:MERLIN}(a). The level of agreement
is similar to the LSWT parametrization shown in Fig.~\ref{fig:MERLIN}(g), with 
the only difference being
that, in the QMC case, the large broadening of the line widths at high energies
makes the intensity dip between the top of the V shape and the high-energy flat
mode less prominent. The disappearance of the flat high-energy mode and
broadening of the V-shaped feature with increasing temperature in the
paramagnetic phase, shown in Figs.~\ref{fig:MERLIN}(a)--\ref{fig:MERLIN}(c), 
also seem to be
qualitatively captured by QMC calculations at the corresponding temperatures,
see Figs.~\ref{fig:MERLIN}(d)--\ref{fig:MERLIN}(f). The best-fit value for the
intraladder exchange $J=7.07$~meV is close to the range of previous estimates
determined from fits to temperature-dependent susceptibility data 
(7.3--8.1~meV). \cite{Gibbs2017, Glamazda2017} The extracted interladder 
exchange $J'=0.58J$ is larger than the theoretically predicted critical value
$J'_\mathrm{c}=0.314J$ for a 2D model of coupled ladders, \cite{Matsumoto2001}
placing \BCTO{} in the N\'eel-ordered state beyond the quantum critical point
(as indicated in Fig.~\ref{fig:QPD}), consistent with the experimental
observations of a finite $T_\mathrm{N}$. \cite{Gibbs2017,Glamazda2017,Rao2016}
We note that the best-fit
value for the exchange $J$ differs between the LSWT and QMC calculations, as the
latter includes the effects of higher-order quantum fluctuations. We therefore
expect the values found from comparison with the QMC calculation to be closer to
the actual exchange parameters in the material.

\subsubsection{Spin dynamics for ladders at criticality}
\label{sec:criticalJ'}

\begin{figure}
\centering
\includegraphics[width=\linewidth,keepaspectratio]{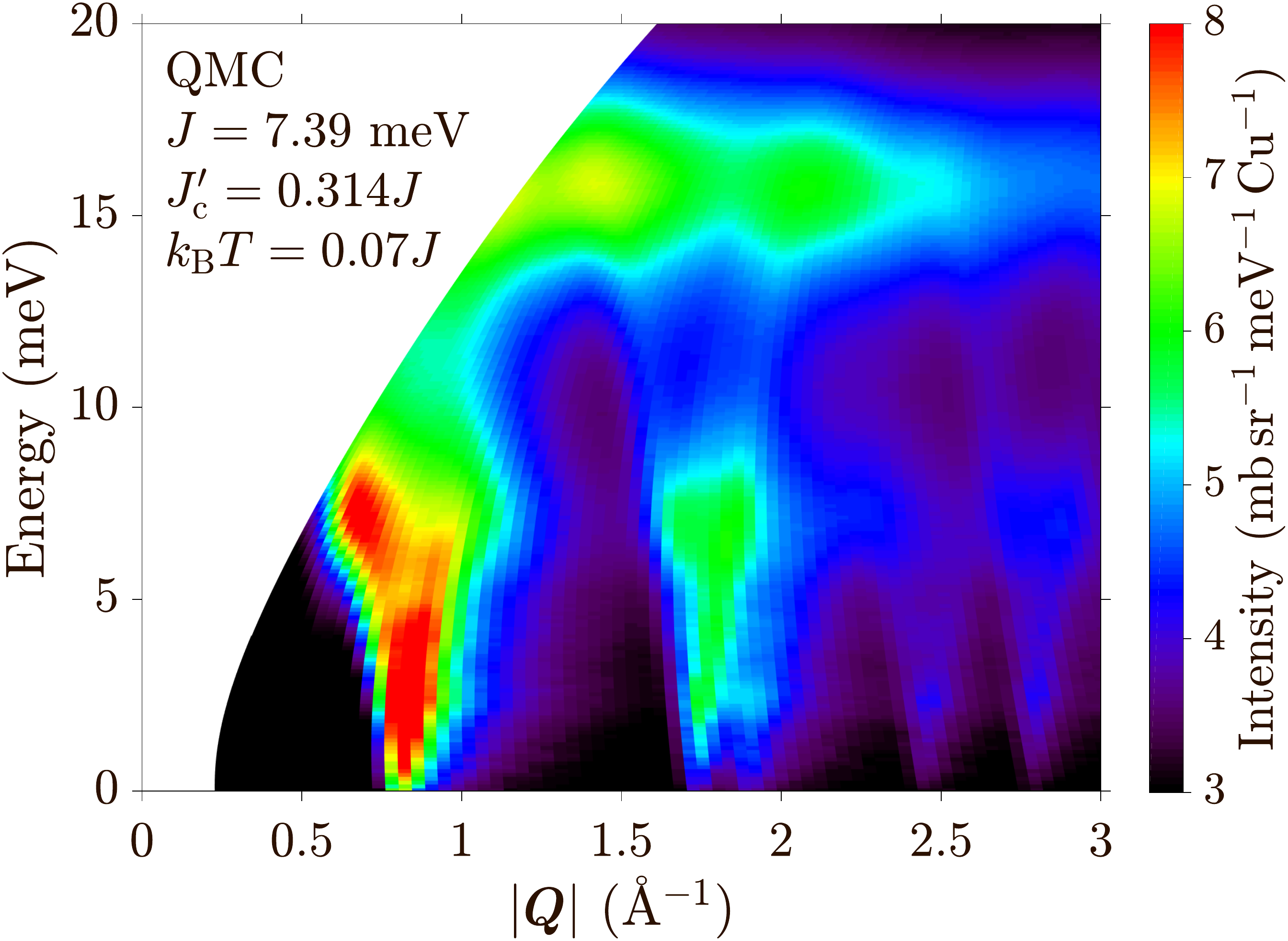}
\caption{(Color online) Spherically averaged spin dynamics for ladders at the
quantum critical coupling $J'_\mathrm{c}=0.314J$, calculated using QMC, to be
contrasted with the measured powder INS data in Fig.~\ref{fig:MERLIN}(a). The
scale $J=7.39$~meV was chosen to best reproduce the location of the flat mode in
the data, and $T$ corresponds to the base temperature of the experiment
$k_\mathrm{B}T=0.07J$.\label{fig:QMC_powder_critJp}}
\end{figure}

For a system of coupled two-leg spin-$\half{1}$ ladders with equal leg and rung
couplings, the quantum critical point occurs at $J'_{\mathrm{c}}=0.314J$.
\cite{Matsumoto2001} For completeness, we show the QMC calculation for this
critical coupling in Fig.~\ref{fig:QMC_powder_critJp}, to be compared with the
data in Fig.~\ref{fig:MERLIN}(a). The value of $J$ in this calculation was
chosen to reproduce the measured energy of the flat mode in the data. A notable
difference compared to the data is that the narrowing at the top of the V-shaped
dispersion, used previously to determine the $J'$ value, is clearly larger than
experimentally observed, confirming that \BCTO{} has a stronger $J'$ and is
therefore located deeper in the ordered side of the phase diagram in
Fig.~\ref{fig:QPD}.

\section{High field magnetization}
\label{sec:magnetisation}

\begin{figure}
\centering
\includegraphics[width=\linewidth,keepaspectratio]{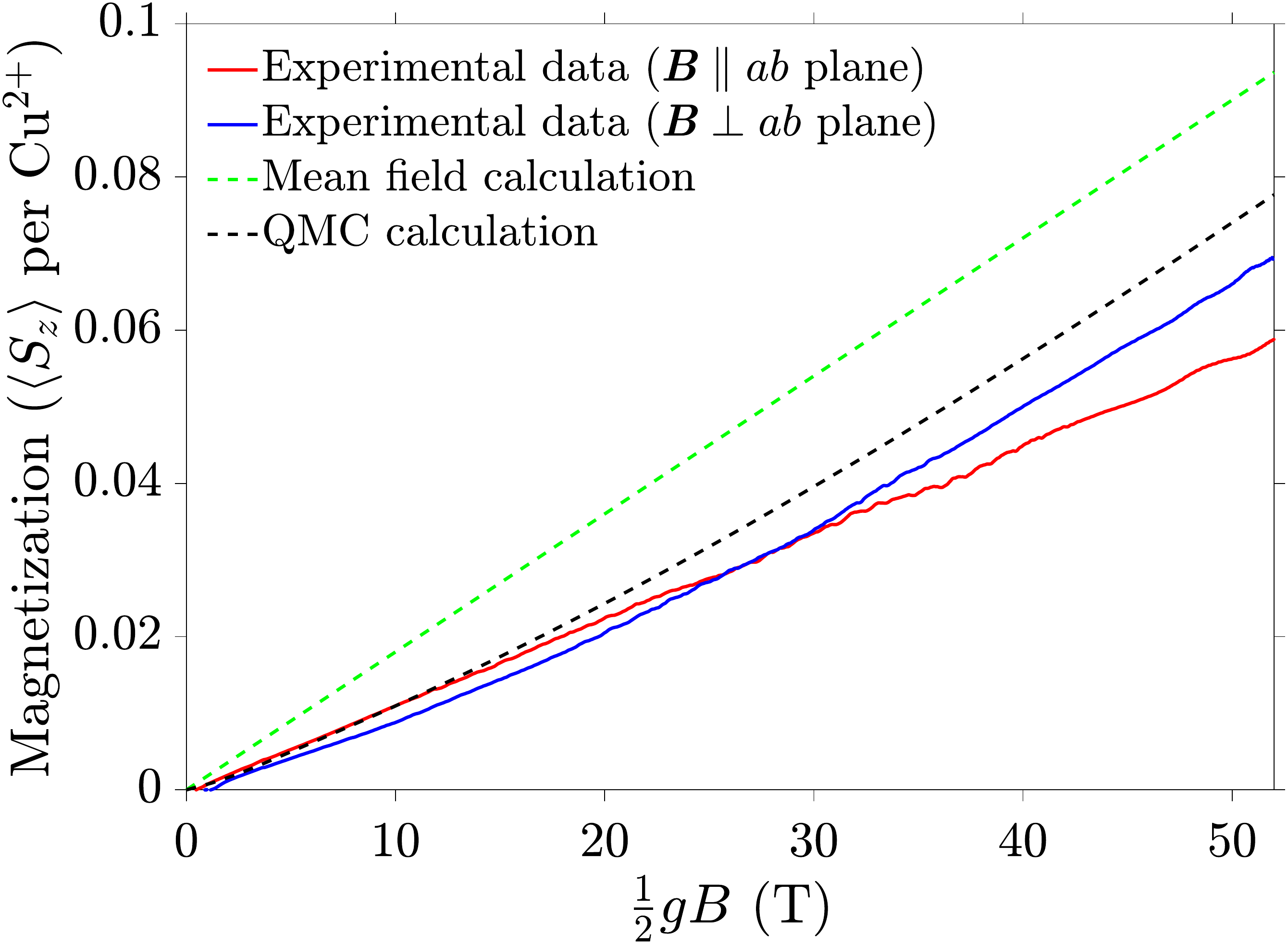}
\caption{(Color online) Longitudinal magnetization curve in single crystals of
\BCTO{} as a function of the applied field, for
$\vect{B}\parallel{}ab$ plane (red line) and $\vect{B}\perp{}ab$ plane (blue
line). \cite{Gibbs2017} The mean-field theory prediction assuming a spin-flop
phase is shown by the green dashed line [exchange parameters as in
Fig.~\ref{fig:LSWT&QMC}(a)] and the QMC calculation by the black dashed line
[exchange parameters as in
Fig.~\ref{fig:LSWT&QMC}(b)].\label{fig:Magnetisation}}
\end{figure}

As a consistency check of the overall energy scale of the interactions deduced
from fits to the LSWT and
QMC calculations, we compare below the magnetization curve as a function of
applied magnetic field with the predictions of the two theoretical models.
Figure~\ref{fig:Magnetisation} shows the previously measured high field
magnetization data for magnetic fields applied parallel and perpendicular to the
$ab$ plane (red and blue solid lines, respectively), where the horizontal scale
is the renormalized magnetic field $\half{1}gB$, assuming
$g_{\parallel{}ab}=2.080$ and $g_{\perp{}ab}=2.289$. \cite{Gibbs2017} The slight
difference between the two curves is consistent with the assumption that the
system is almost isotropic, with only rather small Dzyaloshinskii-Moriya
or other exchange anisotropy terms.
Neglecting such small anisotropies, which
are beyond the scope of the present analysis, the mean-field prediction assuming
a spin-flop phase is shown as the upper dashed green line and the QMC
calculation as a dashed black line. The experimental magnetization data (average
of the two solid lines) is significantly reduced compared to the mean-field
prediction (which neglects entirely zero-point quantum fluctuations) and is
quite close to the QMC calculation. We regard this agreement as a consistency
check of the sum of the exchanges [given by term $A$ in Eq.~(\ref{eq:ACD})]
deduced by comparing the observed spin dynamics with the QMC calculations.

\section{Conclusion}
\label{sec:conclusion}
To summarize, we have reported inelastic neutron scattering measurements of the 
spin dynamics in \BCTO{} over the full bandwidth of magnetic excitations. The 
observed spectrum is consistent with that expected for two-leg antiferromagnetic
ladders with finite interladder couplings in the $bc$ plane and almost
negligible couplings between planes. Through quantitative comparison with both
linear spin wave theory and quantum Monte Carlo calculations, we have proposed
values for all relevant exchange parameters, both intraladder as well as 
interladder. The deduced values put the system on the ordered side of the phase
diagram for coupled two-leg ladders, in proximity to the critical point where 
the magnetic order is suppressed.

\begin{acknowledgments}
D.M. acknowledges support from an EPSRC doctoral studentship. Work in Oxford
was partly supported by the EPSRC Grant No. EP/M020517/1 and the ERC Grant No. 
788814 (EQFT). T.Y. and S.W. acknowledge support by the Deutsche 
Forschungsgemeinschaft (DFG) under Grants FOR 1807 and RTG 1995. Furthermore, 
they thank the IT Center at RWTH Aachen University and the JSC J\"ulich for 
access to computing time through JARA-HPC. T.Y. is also supported by the 
National Natural Science Foundation of China (NSFC Grant No. 11504067). F.M. 
acknowledges the hospitality of the Max Planck Institute for Solid State 
Research in Stuttgart and the financial support of the Swiss National Science 
Foundation (SNF). The neutron scattering measurements at ISIS Neutron and Muon 
Source were supported by a beam time allocation from the Science and Technology 
Facilities Council. In accordance with the EPSRC policy framework on research 
data, access to the data will be made available from 
Ref.~\onlinecite{data_archive}.
\end{acknowledgments}

\appendix

\begin{figure}
\centering
\includegraphics[width=\linewidth,keepaspectratio]{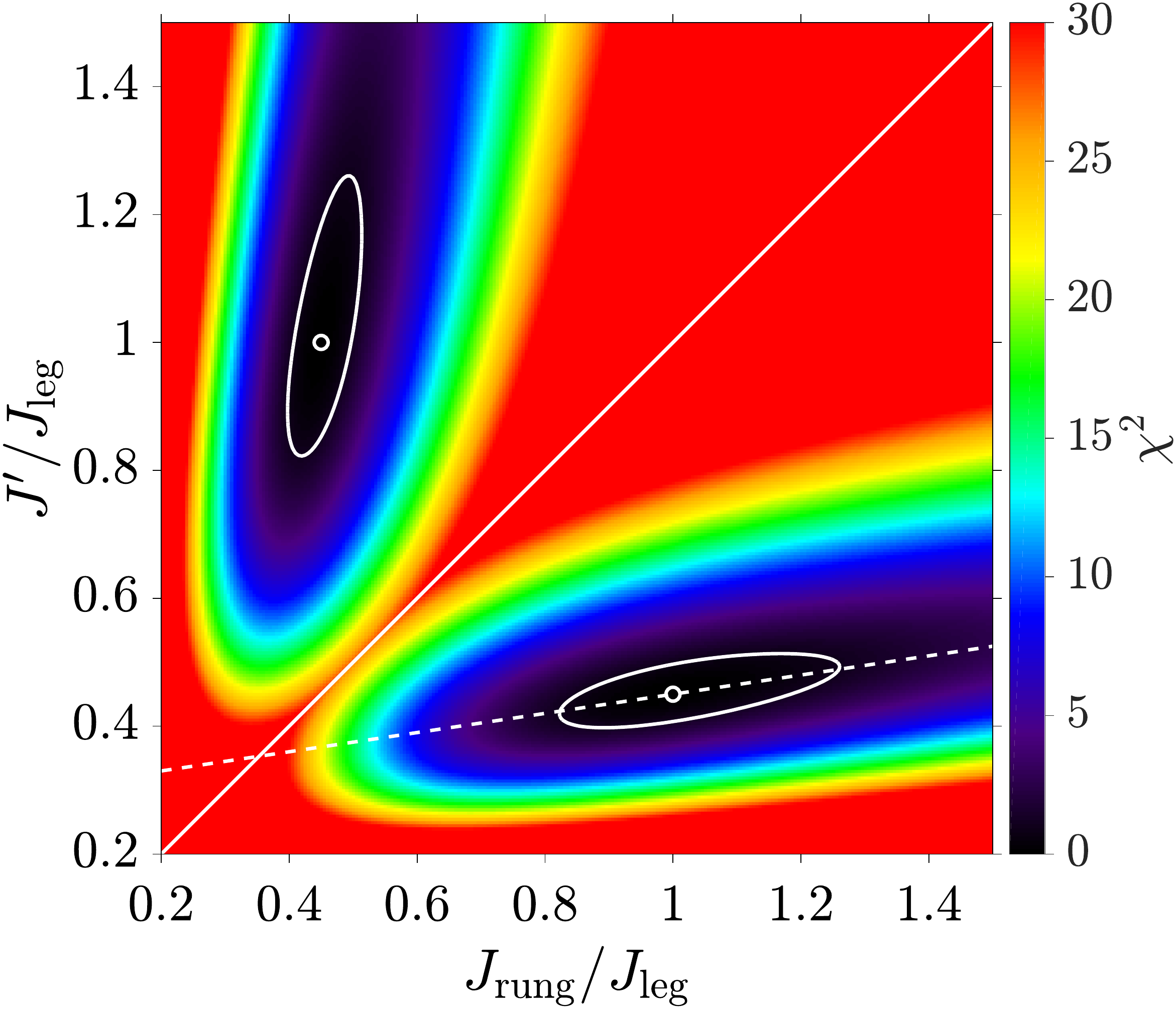}
\caption{(Color online) $\chi^2$ as a function of $J'/J_\mathrm{leg}$ and
$J_\mathrm{rung}/J_\mathrm{leg}$ for the LSWT model of coupled two-leg
spin-$\half{1}$ ladders, quantifying the deviation from the calculation for the
best-fit parameters [see Eq.~(\ref{eq:Chi2})]. The solid white line represents
$J_\mathrm{rung}=J'$. Below this line is the physically expected case of weakly
coupled ladders, as shown in Fig.~\ref{fig:Structure}(b). $J_\mathrm{leg}$ was
chosen to match the flat mode energy to the experimentally observed value. The
parameter values for the fitted model ($J_\mathrm{rung}=J_\mathrm{leg}$ and
$J'=0.45J_\mathrm{leg}$) are indicated by white circles (including the
equivalent model with $J'\leftrightarrow{}J_\mathrm{rung}$). The dashed white
line is a plot of Eq.~(\ref{eq:Jp_Jrung}), discussed in the text. The white
ovals are the $\chi^2=1$ ($1 \sigma$) contours.\label{fig:LSWT_chi2}}
\end{figure}

\section{Linear spin wave theory calculation}
\label{sec:appendix_LSWT}

This section outlines the LSWT calculation (introduced in Sec.~\ref{sec:LSWT})
of the dispersion relation and dynamical structure factor for neutron scattering
from two-leg ladders arranged in planes with monoclinic stacking. Following
\textit{ab initio} calculations, \cite{Gibbs2017} we assume AFM Heisenberg
exchange interactions along the legs ($J_\mathrm{leg}$), along the rungs
($J_\mathrm{rung}$), between the ladders in the $bc$ plane ($J'$), and between
next-nearest-neighbor planes ($J_\mathrm{3D}$). The exchange paths and the
relative spin alignments in the mean-field ground state are shown in
Fig.~\ref{fig:Structure}. We neglect the coupling $J_\mathrm{f}$ between offset
adjacent planes as it is frustrated [see Fig.~\ref{fig:Structure}(c)], leading
to decoupled adjacent ladder planes at the mean-field level. So at this level of
the approximation, the light and dark blue Cu$^{2+}$ sites in
Fig.~\ref{fig:Structure} form two magnetically decoupled subsystems. The
calculation focuses on the dark blue sites with two Cu$^{2+}$ ions per primitive
structural unit cell, labeled 1 and 2 in Figs.~\ref{fig:Structure}(a) and 
\ref{fig:Structure}(b).

To describe the spin axes, we use a Cartesian coordinate system $(x,y,z)$ with
$\vect{\hat{z}}$ along the direction of the ordered moments in the ground state
($b$ axis). The analytic calculation is simplified by performing a rotation of
the local spin axes in the $xz$ plane by an angle
$\alpha(\vect{r})=\vect{Q}_\mathrm{rot}\cdot\vect{r}$, where $\vect{r}$ is the
spin position and $\vect{Q}_\mathrm{rot}=(\half{1},\half{1},0)$. In this
rotating frame, the magnetic unit cell then reduces to the same size as the
primitive structural cell, with two sublattices labeled 1 and 2 in
Figs.~\ref{fig:Structure}(a) and \ref{fig:Structure}(b). Using a 
Holstein-Primakoff transformation, a Fourier transformation, and neglecting 
terms higher than quadratic order, the spin Hamiltonian in this rotated frame is
obtained as
\begin{equation}
\mathcal{H}=\frac{1}{2}\sum_{\vect{k}}\matr{X}^\dagger\matr{H}\matr{X}
-N(S+1)\frac{A}{2},
\label{eq:Hamiltonian}
\end{equation}
where $N$ is the total number of spin sites, and the sum is over all wave 
vectors $\vect{k}$ in the first Brillouin zone of the primitive structural unit
cell. The operator basis is chosen to be $\matr{X}^\dagger=
\begin{pmatrix}
a_{\vect{k}}^\dagger&b_{\vect{k}}^\dagger&a_{-\vect{k}}&b_{-\vect{k}}
\end{pmatrix}$, where $a$ and $b$ refer to the magnetic sublattices numbered 1
and 2 in Figs.~\ref{fig:Structure}(a) and \ref{fig:Structure}(b), such that 
$a_{\vect{k}}^\dagger$
($a_{\vect{k}}$) creates (annihilates) a plane wave magnon on the first
sublattice and likewise for $b$ on the second sublattice. The Hamiltonian matrix
has the form
\begin{equation}
\matr{H}=
\begin{pmatrix}
A&0&C&D^\ast\\
0&A&D&C\\
C&D^\ast&A&0\\
D&C&0&A
\end{pmatrix},
\label{eq:Hamiltonian_matrix}
\end{equation}
where $A$, $C$, and $D$ are given by Eq.~(\ref{eq:ACD}). Using standard methods
to diagonalize the bilinear boson Hamiltonian \cite{White1965} and rotating back
to the fixed laboratory frame gives the dispersion relations listed in
Eq.~(\ref{eq:Dispersion}), which by periodicity hold for a general wave vector
$\vect{Q}$ in reciprocal space. There are two dispersion branches as there are
two sites in the magnetic cell.

The dynamical structure factor (per spin in the laboratory frame) for spin
fluctuations along the $x$ direction is obtained as
\begin{multline}
S^{xx}(\vect{Q},\omega)=\\
\frac{S}{4}(1+\cos{\phi})\frac{A+C-D_0}{\hbar\omega_{\vect{Q}}^+}
[n(\hbar\omega_{\vect{Q}}^+)+1]\mathcal{N}(\hbar\omega_{\vect{Q}}^+,\sigma)\\
+\frac{S}{4}(1-\cos{\phi})\frac{A+C+D_0}{\hbar\omega_{\vect{Q}}^-}
[n(\hbar\omega_{\vect{Q}}^-)+1]\mathcal{N}(\hbar\omega_{\vect{Q}}^-,\sigma),
\label{eq:Sxx}
\end{multline}
where $\mathcal{N}(\hbar\omega,\sigma)$ is a Gaussian function with the center
at $\hbar\omega$ and standard deviation $\sigma$, used to model the instrumental
energy resolution ($\mathrm{FWHM}=2\sqrt{2\ln2}\sigma$). Note that the finite
temperature Bose factor $n(\hbar\omega)+1$, where
$n(\hbar\omega)=1/\left(e^{\hbar\omega/k_\mathrm{B}T}-1\right)$, has been
included in the definition of the dynamical structure factor for consistency
with the notation used in the QMC calculations in
Appendix~\ref{sec:appendix_QMC}. The above analytic expressions for the
dispersion and dynamical structure factor were checked explicitly against the
numerical predictions of \textsc{spinw}. \cite{Toth2015}

The one-magnon neutron scattering cross section, including the polarization
factor and the magnetic form factor, is then
\begin{equation}
I(\vect{Q},\omega)=(\gamma{}r_0)^2\left(1+\frac{Q_z^2}{\abs{\vect{Q}}^2}\right)
\left[\frac{g}{2}f(\abs{\vect{Q}})\right]^2S^{xx}(\vect{Q},\omega),
\label{eq:Intensity}
\end{equation}
where $(\gamma{}r_0)^2=290.6$~mb\,sr$^{-1}$ is a factor that converts the
intensity into absolute units of \units, $f(\abs{\vect{Q}})$ is the spherical
magnetic form factor for Cu$^{2+}$ ions, $Q_z$ is the component of the
wave vector $\vect{Q}$ along the $z$ direction ($b$ axis), and the $g$ factor is
assumed equal to 2. Equation~(\ref{eq:Intensity}) was spherically averaged to
produce Figs.~\ref{fig:MERLIN}(g) and \ref{fig:LET}(d) for direct comparison
with the powder INS data. The spherically averaged INS spectrum is very similar
for different spin directions in the ground state with only slight changes in
intensity modulations, which cannot be reliably differentiated using the
experimental INS data. For concreteness, we have therefore assumed the ordered
moments to be aligned along $\vect{b}\parallel{}\vect{\hat{z}}$ and have used
the corresponding polarization factor in all calculations of the INS intensity.

\section{Unequal leg and rung couplings}
\label{sec:appendix_legrung}

The assumption that $J_\mathrm{leg}=J_\mathrm{rung}$ was tested by comparing the
data to the LSWT model for isolated ladder planes ($J_\mathrm{3D}=0$) with
variable $J'$ and $J_\mathrm{rung}$ relative to $J_\mathrm{leg}$ and calculating
a corresponding goodness of fit $\chi^2$, defined as
\begin{equation}
\chi^2 = \frac{(\hbar\hat{\omega}^+ - \hbar\omega^+)^2}{(\sigma^+)^2} +
\frac{(\hbar\hat{\omega}^- - \hbar\omega^-)^2}{(\sigma^-)^2},
\label{eq:Chi2}
\end{equation}
where $\hbar\omega^+=2S\sqrt{J_\mathrm{rung}(2J_\mathrm{leg}+J')}$ and
$\hbar\omega^-=2S\sqrt{J'(2J_\mathrm{leg}+J_\mathrm{rung})}$ are the energies at
$(0,\half{1},\half{1})$ of the even and odd magnon modes [using
Eq.~(\ref{eq:Dispersion})], respectively, and $\hbar\hat{\omega}^+$ and
$\hbar\hat{\omega}^-$ are these energies for the best-fit parameters
$\hat{J}_\mathrm{leg}=\hat{J}_\mathrm{rung}=9.32$~meV and
$\hat{J}'=0.45\hat{J}_\mathrm{leg}$. For each $(J_\mathrm{rung},J')$ pair,
$J_\mathrm{leg}$ was chosen to keep the energy of the flat mode in the
spherically averaged spectrum fixed at the best-fit value [fixing $A$ in
Eq.~(\ref{eq:ACD})]. $\sigma^+=0.6$~meV and $\sigma^-=0.4$~meV are the
uncertainties in the fitted positions of the even and odd modes at
$(0,\half{1},\half{1})$, respectively, estimated by comparing the energy scan in
Fig.~\ref{fig:MERLIN}(h) to models with variable $\hbar\omega^+$ and
$\hbar\omega^-$ at a constant flat mode energy. $\chi^2$ for a range of
$J'/J_\mathrm{leg}$ and $J_\mathrm{rung}/J_\mathrm{leg}$ values is plotted as a
color map in Fig.~\ref{fig:LSWT_chi2}. The mirror symmetry about
$J_\mathrm{rung}=J'$ (solid white line) is to be expected, as the system is
invariant under interchange of rung and interladder couplings [for
$J'>J_\mathrm{rung}$, the $J_\mathrm{rung}$ exchange acts as an interladder
coupling for ladders with rung coupling $J'$, see Fig.~\ref{fig:Structure}(b)].
Assuming $J'<J_\mathrm{rung}$ and fixing the energy of the flat mode and the
(lower energy) odd mode at $(0,\half{1},\half{1})$, the following relationship
is found between $J'$ and $J_\mathrm{rung}$ using Eq.~(\ref{eq:Dispersion}):
\begin{equation}
J'=\frac{\hat{J}'(2J_\mathrm{leg}+J_\mathrm{rung})}
{2\hat{J}_\mathrm{leg}+\hat{J}_\mathrm{rung}},
\label{eq:Jp_Jrung}
\end{equation}
which is plotted as a dashed white line in Fig.~\ref{fig:LSWT_chi2}. Starting
from the best-fit parameters (white circle) and moving along the line with
increasing $J_\mathrm{rung}/J_\mathrm{leg}$, the even mode at
$(0,\half{1},\half{1})$ increases in energy and the gap in signal near 13~meV in
Fig.~\ref{fig:MERLIN}(h) widens. If $J_\mathrm{rung}/J_\mathrm{leg}$ is
decreased from the fitted value, the even mode decreases in energy and the gap
between the even and odd modes [shown in Fig.~\ref{fig:LSWT&QMC}(a)] closes and
disappears on the $J_\mathrm{rung}=J'$ line, at which point the system consists
of a rectangular lattice of anisotropic couplings ($J$ along $\vect{b}$ and $J'$
along $\vect{c}$). The $\chi^2=1$ contour lines in Fig.~\ref{fig:LSWT_chi2}
(white ovals), which correspond to a $1\sigma$ deviation from the best-fit
parameter values, show that there is a finite range of
$J_\mathrm{rung}/J_\mathrm{leg}$ values for which the LSWT model could provide a
good fit to the INS data. This suggests that the ratio of rung and leg couplings
$J_\mathrm{rung}/J_\mathrm{leg}$ is likely to be between 0.8 (weaker rungs) and
1.3 (stronger rungs), with $J'/J_\mathrm{leg}$ adjusted accordingly ($J'$
reduced on the weak rung side).

\section{Quantum Monte Carlo calculation}
\label{sec:appendix_QMC}

\begin{figure}
\centering
\includegraphics[width=\linewidth,keepaspectratio]{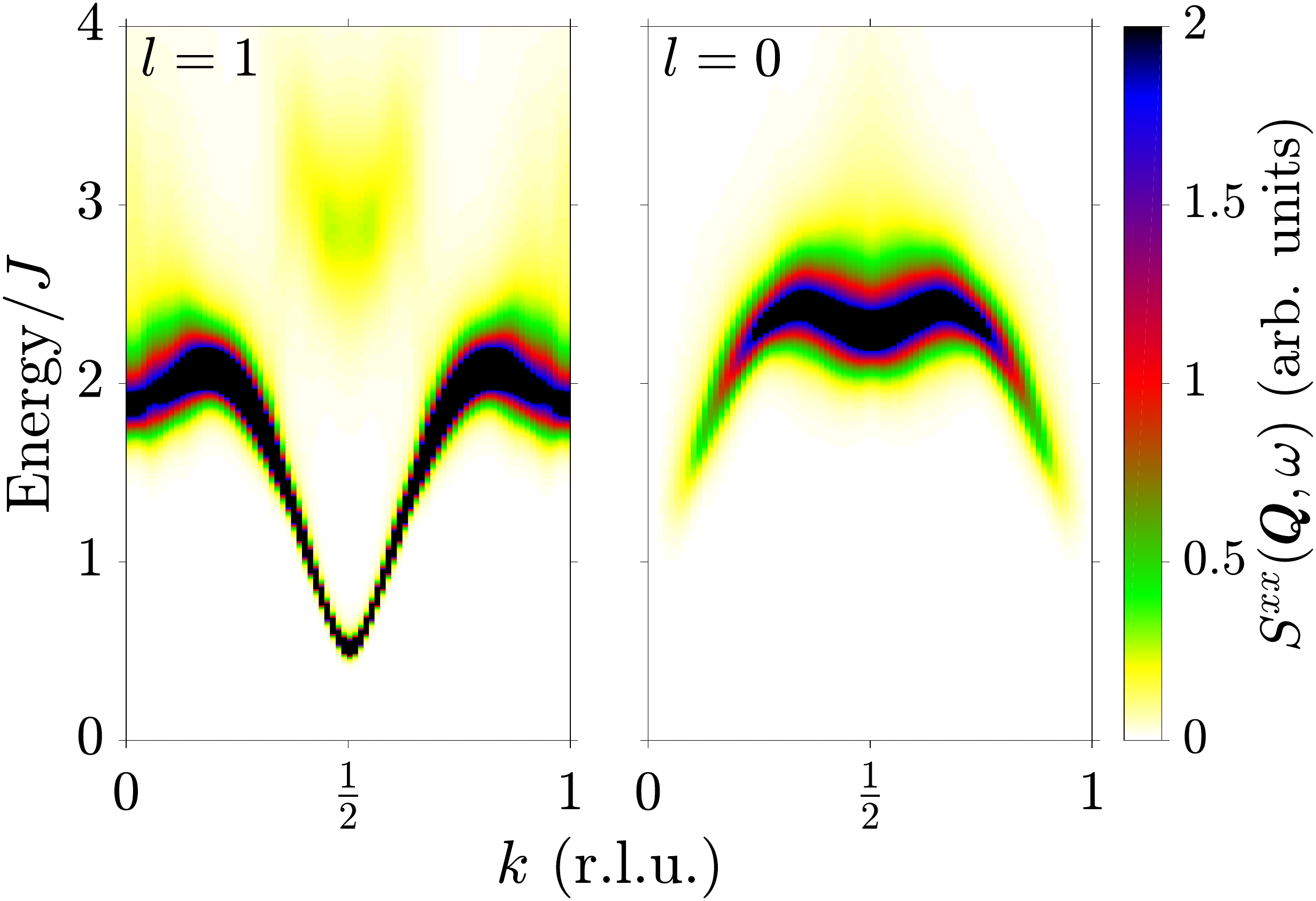}
\caption{(Color online) QMC calculation of the dynamical spin structure factor
$S^{xx}(\vect{Q},\omega)$ along the ladder direction for a single isolated
two-leg spin-$\half{1}$ ladder with equal leg and rung couplings $J$ at
temperature $k_\mathrm{B}T = 0.07J$. The left-hand (right-hand) panel shows the
odd (even) channel, cf. the coupled ladders calculation in the second (fourth)
panel of Fig.~\ref{fig:LSWT&QMC}(b) (the interladder dispersions plotted in the
first and third panels would be flat for an isolated ladder).
\label{fig:QMC_decoupled}}
\end{figure}

We performed QMC simulations for a plane of parallel ladders, as in
Fig.~\ref{fig:Structure}(b), described by the Hamiltonian
\begin {equation}\label{eq:Ham}
\mathcal{H} = J \sum_{\langle i,j \rangle} \vect{S}_i \cdot \vect{S}_j + J'
\sum_{\langle i,j \rangle'} \vect{S}_i \cdot \vect{S}_j,
\end{equation}
where $J$ denotes the coupling within the ladders (taken equal along legs and
rungs) and $J'$ the coupling between neighboring ladders. We used the stochastic
series expansion method with directed loop updates. \cite{Sandvik1999,
Syljuasen2002, Alet2005} For the QMC simulations, a Cartesian coordinate system
$(x,y)$ was defined such that $\vect{\hat{x}}$ is aligned parallel to the ladder
(leg) direction and $\vect{\hat{y}}$ is the perpendicular direction (parallel to
the ladder rungs). We assumed the ladders are equally spaced along the $y$
direction and entirely confined to the 2D plane, which corresponds to
Fig.~\ref{fig:Structure}(b) with $\vect{\hat{x}}$ along $\vect{b}$,
$\vect{\hat{y}}$ along $\vect{c}$, and spin sites confined to the $bc$ plane
($\xi=0$) and equally spaced along $\vect{c}$ ($\zeta=\half{1}$). We considered
a finite system with $N_\mathrm{s}=40\times 40$ spins and used periodic boundary
conditions in both lattice directions. The dynamical spin structure factor is
defined as
\begin{equation*}\label{eq:Sfactor}
S(\vect{Q}, \omega) = \frac{1} {N_\mathrm{s}}\int{dt}{\sum_{j,k}e^{i[\omega
t-\vect{Q} \cdot (\vect{r}_j-\vect{r}_k)]}\langle \vect{S}_j(t) \cdot
\vect{S}_k(0)\rangle}.
\end{equation*}
For the QMC simulations, it is convenient to express the dynamical spin
structure factor in an explicit unit cell decomposition for which each unit cell
contains one rung of the coupled ladders. In the following, $\mu$ and $\nu$
denote such unit cells and $\vect{S}_{\mu1}$ ($\vect{S}_{\mu2}$) denotes the
lower (upper) spin in the $\mu$th unit cell. Furthermore, we set the position
vector of the spins such that $\vect{r}_{\mu1}=\vect{R}_\mu$ and
$\vect{r}_{\mu2}=\vect{R}_\mu+\vect{\delta}$. Here, $\vect{\delta}=(0,
\half{1})$ denotes the vector connecting the two spins within a unit cell, and
$\vect{R}_\mu$ the position vector of the $\mu$th unit cell. The number of spins
$N_\mathrm{s}$ and the number of unit cells $N_\mathrm{u}$ are related by
$N_\mathrm{s}=2 N_\mathrm{u}$. We then obtain
\begin{equation*}
S(\vect{Q},\omega) =
\cos^2{\frac{\vect{Q}\cdot\vect{\delta}}{2}}S^+(\vect{Q},\omega) +
\sin^2{\frac{\vect{Q}\cdot\vect{\delta}}{2}}S^-(\vect{Q},\omega)
\end{equation*}
in terms of the even and odd structure factors with respect to the ladder
reflection symmetry,
\begin{multline*}
S^{\pm}(\vect{Q},\omega) = \frac{1} {2
N_\mathrm{u}}\int{dt}{\sum_{\mu,\nu}}e^{i[\omega t-\vect{Q} \cdot
(\vect{R}_\mu-\vect{R}_\nu)]}\\
\times\langle [\vect{S}_{\mu1}(t) \pm \vect{S}_{\mu2}(t) ]  \cdot
[\vect{S}_{\nu1}(0)\pm \vect{S}_{\nu2}(0)  ] \rangle,
\label{eq:SfactorU}
\end{multline*}
which are more conveniently obtained separately in the QMC simulations. The
calculations were performed using an efficient scheme to measure imaginary time
displaced spin-spin correlation functions. \cite{Michel2007a,*Michel2007} The
dynamical spin structure factor was then obtained after an analytic continuation
based on the stochastic formulation of Ref.~\onlinecite{Beach2004}. The powder
spectrum was finally obtained from the QMC dynamical spin structure factor
$S(\vect{Q},\omega)$ by applying the same spherical averaging procedure as for
the LSWT model.

Due to the statistical noise, the analytic continuation broadens the spectral
functions, in addition to any intrinsic and thermal broadening. The spectra for
the QMC model therefore exhibit an enhanced broadening compared to the LSWT
model, see Fig.~\ref{fig:LSWT&QMC}. However, the observed broadening is further
enhanced within the high-energy region around $E\simeq16$~meV. As discussed in
Sec.~\ref{sec:2magnons}, multimagnon scattering processes may lead to a lifetime
broadening of the single-magnon modes in this energy range. Since it is
difficult for the analytic continuation scheme to separate such broadened magnon
modes from the multimagnon continuum contributions, we obtain a broadened QMC
spectrum at these energies. We also observed such enhanced spectral broadening
in the high-energy range for the QMC dynamical spin structure factor of a single
isolated ladder ($J'=0$), see Fig.~\ref{fig:QMC_decoupled}, which can be
compared to previous calculations based on the density matrix renormalization
group (DMRG) approach. \cite{Schmidiger2013} In the even channel of the two-leg
ladder, a weak two-magnon continuum is located close to a spin-1 bound state
within this energy range. Since it is difficult for analytic continuation
methods to separate the two contributions, the QMC signal is broadened in this
channel. The single-magnon mode in the odd channel of the two-leg ladder is
close to the multimagnon continuum at small momenta, which leads to similarly
enhanced broadening. It may be worthwhile for future research to examine in more
detail the evolution of the spectral function within this elevated energy range
as a function of the interladder coupling strength, connecting these
single-ladder results to the strongly coupled case.

\section{Monoclinic--triclinic unit cell transformation}
\label{sec:appendix_transformation}

\begin{figure}
\centering
\includegraphics[width=\linewidth,keepaspectratio]{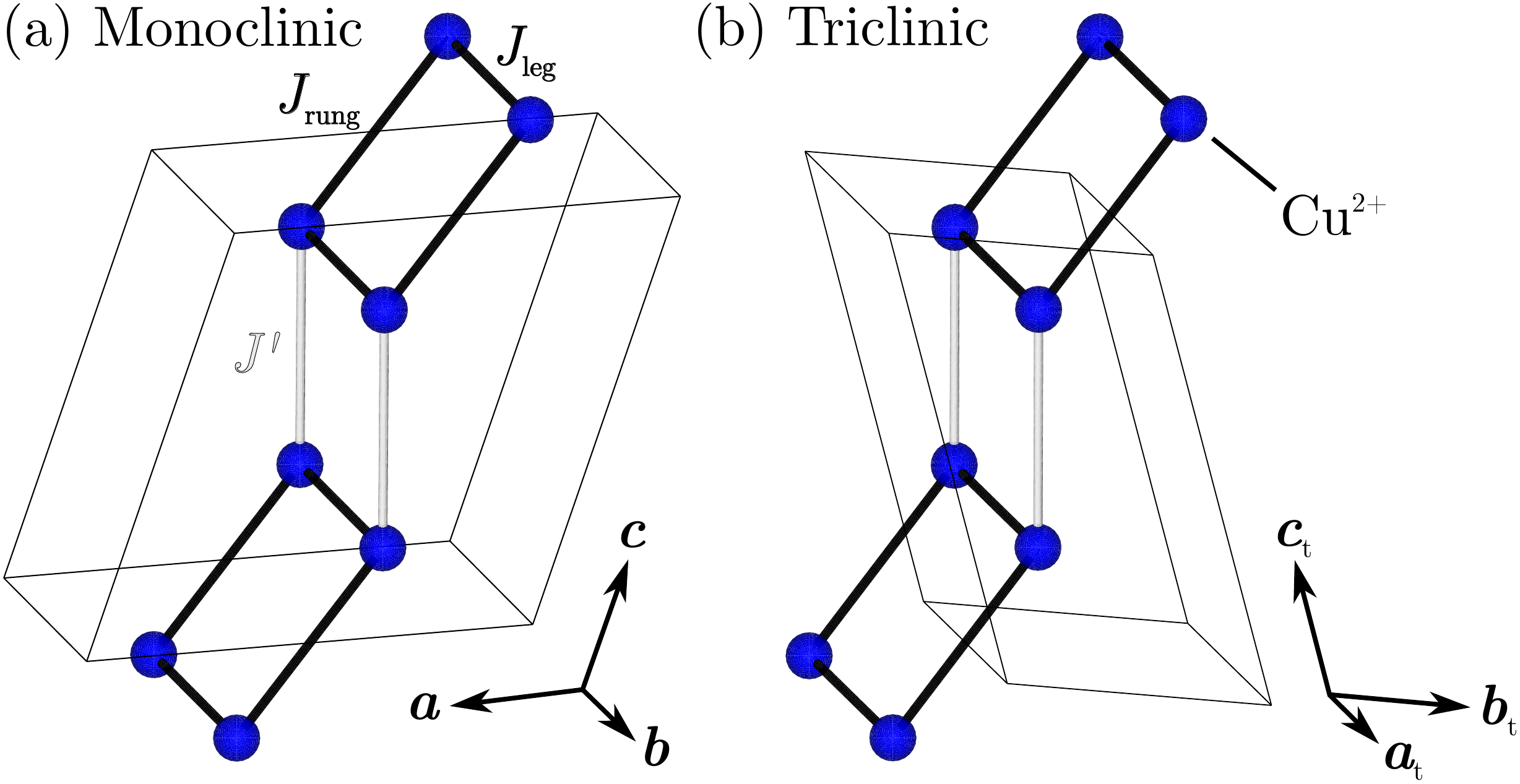}
\caption{(Color online) Section of two coupled ladders in \BCTO{}, showing (a)
the monoclinic and (b) the triclinic unit cells (thin black outlines). The
diagrams include the positions of the Cu$^{2+}$ ions (blue circles), the
intraladder couplings $J_\mathrm{leg}$ and $J_\mathrm{rung}$ (thick black
lines), and the interladder coupling $J'$ along $\vect{c}$ (gray lines). The
diagrams were produced using \textsc{vesta}.
\cite{Momma2011}\label{fig:Unit_cells}}
\end{figure}

The room-temperature crystal structure of \BCTO{} is monoclinic ($C2/m$), and a
weak structural distortion to a triclinic phase ($P\bar{1}$) occurs at
$T_\mathrm{S}=287$~K. \cite{Gibbs2017,Glamazda2017} We use throughout the higher
symmetry monoclinic unit cell description, since the triclinic distortion is
very small. The monoclinic lattice parameters are $a=10.2444(3)$~\AA,
$b=5.7315(2)$~\AA, $c=10.1055(5)$~\AA, and $\beta=108.019(3)^\circ$ at
$T=296$~K. \cite{Gibbs2017} Ignoring this small distortion, the transformation
from the monoclinic lattice basis vectors ($\vect{a}$, $\vect{b}$, $\vect{c}$)
to the triclinic ones ($\vect{a}_\mathrm{t}$, $\vect{b}_\mathrm{t}$,
$\vect{c}_\mathrm{t}$) is given by
\begin{equation}
\begin{pmatrix}
\vect{a}_\mathrm{t}\\
\vect{b}_\mathrm{t}\\
\vect{c}_\mathrm{t}
\end{pmatrix}
=
\begin{pmatrix}
0&1&0\\
-\frac{1}{2}&\frac{1}{2}&0\\
\frac{1}{2}&-\frac{1}{2}&1
\end{pmatrix}
\begin{pmatrix}
\vect{a}\\
\vect{b}\\
\vect{c}
\end{pmatrix}
.
\label{eq:Transformation}
\end{equation}
The corresponding transformation of the reciprocal lattice vectors is given by
\begin{equation}
\begin{pmatrix}
\vect{a}^\ast_\mathrm{t}\\
\vect{b}^\ast_\mathrm{t}\\
\vect{c}^\ast_\mathrm{t}
\end{pmatrix}
=
\begin{pmatrix}
1&1&0\\
-2&0&1\\
0&0&1
\end{pmatrix}
\begin{pmatrix}
\vect{a}^\ast\\
\vect{b}^\ast\\
\vect{c}^\ast
\end{pmatrix}
,
\label{eq:Transformation_rlv}
\end{equation}
and the wave vector coordinates in reciprocal lattice units transform as
\begin{equation}
\begin{pmatrix}
h_\mathrm{t}\\
k_\mathrm{t}\\
l_\mathrm{t}
\end{pmatrix}
=
\begin{pmatrix}
0&1&0\\
-\frac{1}{2}&\frac{1}{2}&0\\
\frac{1}{2}&-\frac{1}{2}&1
\end{pmatrix}
\begin{pmatrix}
h\\
k\\
l
\end{pmatrix}
,
\label{eq:Transformation_hkl}
\end{equation}
where the subscript ``t'' refers to the triclinic case. The monoclinic and 
triclinic unit cells are shown in Fig.~\ref{fig:Unit_cells} as outlines over a 
section of two coupled ladders.

\end{document}